\documentstyle[11pt,amssymb,amsfonts,amsthm,amssymb,amscd,amstext,amsmath,epsfig]{article}

\def\ben{\begin{equation}}
\def\een{\end{equation}}

    \let\e=\varepsilon

\let\s=\sigma \let\t=\tau

\let\C=\Chi

\def\be{\begin{equation}}
\def\ee{\end{equation}}
\def\ba{\begin{array}}
\def\ea{\end{array}}

\def\ie{\rm i.e.\ }
\def\eg{\rm e.g.\ }
\def\dalemb#1#2{{\vbox{\hrule height .#2pt
        \hbox{\vrule width.#2pt height#1pt \kern#1pt
                \vrule width.#2pt}
        \hrule height.#2pt}}}

\newcommand{\bea}{\begin{eqnarray}}
\newcommand{\eea}{\end{eqnarray}}

\def\ft#1#2{{\textstyle{{\scriptstyle #1}\over {\scriptstyle #2}}}}

\def\R{{{\Bbb R}}}
\def\C{{{\Bbb C}}}

\def\bP{{{\Bbb C}{\Bbb P}}}
\def\bbP{{{\Bbb P}}}
\def\CP{{{\Bbb C}{\Bbb P}}}
\def\RP{{{\Bbb R}{\Bbb P}}}
\def\Z{{{\Bbb Z}}}

\def\cO{{\cal O}}

\begin{document}

\begin{flushright}
DAMTP-2004-141\\
hep-th/0412044
\end{flushright}

\begin{center}
\vspace{2cm}
{ \LARGE {\bf Spacetime foam in twistor string theory}}

\vspace{1.5cm}

Sean A. Hartnoll$^{\dagger}$ and Giuseppe Policastro$^{\sharp}$

\vspace{0.8cm}

$^{\dagger}${\it DAMTP, Centre for Mathematical Sciences,
 Cambridge University\\ Wilberforce Road, Cambridge CB3 OWA, UK}

\vspace{0.3cm}

$^{\sharp}${\it Ludwig-Maximilians-Universit\"at, Theoretische Physik\\
  Theresienstr. 37, 80333 M\"unchen, Germany}

\vspace{0.3cm}

s.a.hartnoll@damtp.cam.ac.uk \hspace{1cm}
policast@theorie.physik.uni-muenchen.de

\vspace{2cm}

\end{center}

\begin{abstract}

We show how a K\"ahler spacetime foam in four dimensional
conformal (super)gravity may be mapped to twistor spaces carrying
the D1 brane charge of the B model topological string theory. The
spacetime foam is obtained by blowing up an arbitrary number of
points in $\C^2$ and can be interpreted as a sum over
gravitational instantons. Some twistor spaces for blowups of $\C^2$
are known explicitly. In these cases we write
down a meromorphic volume form and suggest a relation to
a holomorphic superform on a corresponding super Calabi-Yau manifold.

\end{abstract}

\pagebreak

\tableofcontents

\section{Introduction}\label{Intro}

The fate of spacetime at small length scales remains an
outstanding problem in gravitational physics. The expectation of
severe fluctuations in geometry and topology at the Planck length
has lead to a picture of spacetime foam
\cite{Hawking:zw,wheeler}. However, it is difficult to reliably
quantify Planck
scale physics using general relativity or superstring theory. In
particular, the semiclassical approximation which underpins much
of our knowledge of quantum gravity is unlikely to be valid at
such small scales.

It is therefore rather impressive that the full perturbative
partition function
of K\"ahler gravity \cite{Bershadsky:1994sr}, a topological
gravity theory in six dimensions, has recently been shown to be
expressible as a sum over blowups of an asymptotically fixed toric
Calabi-Yau manifold \cite{Iqbal:2003ds}. This explicit realisation
of spacetime foam was possible because K\"ahler gravity is the
spacetime theory of topological A model string theory
\cite{Witten:hr}. The topological A model on toric Calabi-Yau
manifolds is under good computational control
\cite{Aganagic:2003db,Okounkov:2003sp}.

Combining the A model picture with two other recent developments
in topological string theory suggests an approach to spacetime
foam in a four dimensional theory of gravity. Firstly, it has been
conjectured that ${\mathcal{N}}=4$ conformal supergravity in four
dimensions has a dual description as the topological B model
string theory on ${\mathbb{CP}}^{3|4}$
\cite{Witten:2003nn,Berkovits:2004jj}. These beautiful works have
combined the richness of string theory with the mathematical
structure of twistor theory \cite{Penrose:js,Atiyah:wi}. In
conformal gravities and supergravities the Einstein-Hilbert
action is replaced by
\be
S_{\text{CG}} \propto \int {\rm d}\Omega\, C_{abcd} C^{abcd} \,,
\ee
where $C_{abcd}$ is the Weyl tensor. This action is formally
similar to the Yang-Mills action. Like the Yang-Mills action it is
conformally invariant and power-counting renormalisable. Despite
these attractive features, such theories are generally not thought
to provide adequate theories of gravity because the action
$S_{CG}$ has a four-derivative kinetic term and consequently
suffers from unitarity problems \cite{Fradkin:am}. In a Euclidean
context this is manifested as the inner product not being positive
definite. However, a duality with topological string theory may
help to resolve this issue or at least understand it better.

Secondly, it has been conjectured that there exists an S duality
in topological string theory that relates the A model and B model
on a given Calabi-Yau manifold
\cite{Nekrasov:2004js,Neitzke:2004pf}. The spacetime foam we
mentioned above is due to the worldsheet instantons of the A
model. Worldsheet instantons are S dual to D1 branes, so one
therefore expects that the dual B model spacetime foam will be
described as a gas of D1 branes. In the context of twistor string
theory, a D1 brane wraps a $\bP^1$ in ${\mathbb{CP}}^{3|4}$. Under
the twistor correspondence, each $\bP^1$ in ${\mathbb{CP}}^{3|4}$
corresponds to a point in $\R^4 = \C^2$. A simple operation
that can be performed at a given point is a blowup, that is,
replacing the point with a finite $\bP^1$. This leads to the
suggestion \cite{Nekrasov:2004js} that spacetime foam in conformal
supergravity is described as a sum over blowups, with a twistorial
description given as a gas of D1 branes in the topological B
model. This idea is consistent with the fact that
\cite{Berkovits:2004jj} the perturbative states of the B model are
given by the complex structure deformations of the target space. When
the target space is a twistor space these deformations are mapped to
fluctuations of the metric, \ie perturbative states of
gravity. Considering D1 branes then means extending this
correspondence to a nonperturbative sector.
The objective of this work is to start to flesh out this
correspondence.

There is a further mathematical fact that supports the notion of
spacetime foam in conformal gravity as a sum over blowups of
$\C^2$. This comes from considering gravitational instantons in
conformal gravity. Again like the Yang-Mills action, the action
$S_{\text{CG}}$ is minimised by self-dual or anti-self-dual
instantons with
\be\label{eq:selfdual}
C_{abcd} = \pm \ft{1}{2} \e_{ab}{}^{ef} C_{efcd} \,.
\ee
A metric with a Weyl tensor satisfying (\ref{eq:selfdual}) is
called conformally (anti)-self-dual. A manifold is called
conformally (anti)-self-dual if it admits such a metric. The
twistor correspondence \cite{Penrose:js,Atiyah:wi} states that
there is a one-to-one correspondence between conformally
(anti)-self-dual manifolds and a certain class of three
complex-dimensional manifolds with a real structure.

Given a suitable normalisation, the action evaluated for a
gravitational instanton $M$ is a topological charge.
The charge is the Hirzebruch signature of the four
dimensional manifold
\be
S_{\text{CG}}[M] \propto \frac{1}{96\pi^2}
\int {\rm d}\Omega\, \e^{abef} C_{abcd} C_{ef}{}^{cd}
= \t = b_2^{+} - b_2^{-} \,,
\ee
where $b_2^\pm$ are the second Betti numbers of the manifold.

Blowing up a point of a four manifold $M$ is topologically
equivalent to performing a connected sum with $\overline{\bP^2}$.
Partially because of this fact, and following much of the
mathematical literature, we shall work with anti-self-dual rather
than self-dual manifolds. This choice is related to a choice of
orientation\footnote{$\bP^2$ and $\overline{\bP^2}$ are trivially
homeomorphic, but the homeomorphism is not orientation
preserving.}. Thus we write
\be
M \mapsto M^\prime = M \# \overline{\bP^2} \,.
\ee
We then have that performing the connected sum with
$\overline{\bP^2}$ decreases the signature by one
\be
\t(M^{\prime}) = \t(M) - 1\,.
\ee
A particular case of this is $M^{\prime} = \C^2\#\overline{\bP^2}$
which has $\t(M^{\prime})=-1$. Furthermore this $M^{\prime}$ is a
conformally anti-self-dual manifold and therefore could be thought
of as a `minimal' gravitational instanton. In general $M^{\prime}$
is not conformally anti-self-dual. There are however various deep
results in the theory of anti-self-dual manifolds concerning
direct sums with $\overline{\bP^2}$
\cite{taubes,donaldsonfriedman,lebrunrev1,lebrunrev2}. Amongst
these is the statement that
\be
\begin{array}{c}
\qquad \qquad n \\
M = \C^2\#\overline{\bP^2}\; {\overbrace{ \# \cdots \#}}\;
\overline{\bP^2}\,,
\end{array}
\ee
is conformally anti-self-dual for any $n$. These particular
manifolds have various technically pleasant properties. With some
assumptions about the symmetry of the configurations of
points of $\C^2$ which are
blown up, explicit asymptotically flat anti-self-dual metrics are
known, together with their twistor spaces
\cite{lebrun,lebrunbimer}. These twistor spaces turn out to be
bimeromorphic to projective varieties \cite{lebrun,campana} and
the anti-self-dual metrics turn out to be K\"ahler
\cite{lebrunbimer}.

The upshot of these mathematical remarks is that the sum over
blowups can be thought of as a gas of $\C^2\#\overline{\bP^2}$
gravitational instantons. We thus reach the appealing picture that
an instanton gas in conformal supergravity could correspond to the
D1 brane gas in twistor string theory.

The idea of the vacuum of conformal gravity as a gas of
gravitation instantons was explored in \cite{Strominger:1983ns}
following observations in \cite{Strominger:1984zy}. That work
considered the physics of instantons other than blowups of $\C^2$.
Specifically, the paper studied the instantons $K3$ and $T^4$, as
well as the manifold $S^2\times S^2$ which solves the Euclidean
equations of motion, although it is not conformally anti-self-dual
and therefore not a minimum of the action. These other instantons
should also have an interpretation in twistor string theory that
would be interesting to understand.

In section \ref{corresp} we review the twistor correspondence between
conformally anti-self-dual four manifolds and complex three-folds.
We also review an argument of LeBrun that constructs a natural
super Calabi-Yau manifold from a twistor space and comment on the
topology of twistor spaces.

In section \ref{kahler} we recall some properties of the twistor spaces of
K\"ahler anti-self-dual manifolds. We use these properties to show
that K\"ahler blowups in four dimensions are mapped via the
Penrose transform to a D1 brane charge in twistor space.

In section \ref{blowup} we recall an explicit construction of twistor spaces
for $\C\#\overline{\bP^2}\; \# \cdots \#\; \overline{\bP^2}$, also due
to LeBrun. We discuss 
various aspects of the geometry and topology of these
spaces. Following the prescription developed in section 3, we write
down a meromorphic 3-form that may be integrated to give a D1
brane charge. The form is singular because the canonical bundle of the
twistor space is nontrivial.
We go on to consider a super Calabi-Yau extension of the
twistor spaces and write down a global $(3|4)$-form.
By integrating out the fermionic directions we recover the meromorphic
form on the manifold.

Section \ref{concl} contains our conclusions and a discussion.
The appendices contain technical details of some
of the computations done in the text, as well as a
discussion of geometric transitions in the twistor space.

\section{Twistor correspondence and super Calabi-Yau manifolds}\label{corresp}
 
\subsection{Twistor spaces}\label{twistsp}

Let us briefly review Penrose's construction of a complex
three-manifold $(Z,J)$ from a four dimensional conformally
anti-self-dual Riemannian manifold $(M,[g])$
\cite{Penrose:js,Atiyah:wi}. The manifold $Z$ is called the
twistor space of $M$.

The Hodge * operation acting on 2-forms on $M$ defines a linear map
$\Lambda^2 T^*M \to \Lambda^2 T^*M$ such that $*^2 = 1$.
The eigenspaces corresponding to
eigenvalues $\pm 1$ are the self-dual and anti-self-dual 2-forms
on $M$, respectively. The consequent split $\Lambda^2 = \Lambda^+
\oplus \Lambda^-$ corresponds
to the factorization of the rotation group on the tangent space $SO(4)
\cong SO(3) \times SO(3)$.
We therefore have globally defined vector bundles $\Lambda^\pm$ of
rank 3 with structure group $SO(3)$. The Riemannian metric on $M$
induces a metric on the fibres, so we can consider the bundle
having the unit sphere in $\Lambda^+$ as fibre. The total space of
this $S^2$ bundle is the twistor space $Z = S(\Lambda^+)$ as a
Riemannian 6-manifold.

It is convenient to have a different interpretation for the
fibres. A normalised self-dual 2-form can be identified, using the
metric, with an endomorphism $J: TM \to TM$ that is skew, $J^* =
-J$, and further satisfies $J^2=-1$. A point $z \in Z$ on the
fibre of $x \in M$ is then an almost complex structure $J_x$ on
$T_x M$. Thus the $S^2$ bundle of self-dual 2-forms is also the
bundle of complex structures over $M$. The tangent space of $Z$
splits into $TZ = TF \oplus TM$, where $F$ is the fibre. We can
put an almost complex structure $J$ on $T_z Z$ by firstly defining
it to be $J_x$ on $T_x M$ and then on the fibre taking the unique
complex structure on $S^2 \cong \bP^1$ (up to a sign, chosen to be
compatible with the orientation). The central result is that this
almost complex structure is integrable when the the self-dual part
of the Weyl tensor on $M$ vanishes, $C^+=0$. The twistor space of
an anti-self-dual manifold is therefore a complex 3-fold. This
statement and its converse is given in the following theorem.

\vspace{0.2cm}
\noindent
{\bf Theorem \cite{Penrose:js,Atiyah:wi}:} {\it The almost complex
  twistor space
$(Z,J)$ of $(M,[g])$ is a complex 3-manifold if and only if
$C^+=0$. Conversely, a complex 3-manifold arises by this
construction if and only if it admits an anti-holomorphic
involution $\sigma : Z
\to Z$ without fixed points and a foliation by $\sigma$-invariant
rational curves $\bP^1$, each of which has normal bundle
${\mathcal{O}}(1)\oplus{\mathcal{O}}(1)$}.

\vspace{0.2cm}
The $\sigma$-invariant curves in the twistor space $Z$, called the
real curves, correspond to points in the four-manifold $M$. Two
simple examples of twistor spaces are as follows. The twistor
space of $S^4$ with the conformal equivalence class of the round
metric is $\bP^3$. The twistor space of $\overline{\bP^2}$ with
the Fubini-Study metric is a flag manifold that may be described
as the hypersurface $v
\cdot w = 0$ with $v,w \in \C^3$ homogeneous coordinates on
$\bP^2\times\bP^2$.

The twistor space of $\R^4 = \C^2$ is obtained by removing a
rational curve from $\CP^3$, corresponding to removing a point
from $S^4$ and conformally decompactifying. The resulting space is
written $\CP^{\prime 3}$. We will not always indicate the removal
of the rational curve explicitly.

The construction of $Z$ gives enough information to compute its
Betti numbers using theorems on sphere bundles (\cite{bott}, II,
$\S$ 11). By definition $Z$ is an orientable $S^2$ bundle over
$M$, with associated vector bundle $\Lambda^+$ over $M$. On an
$S^n$-bundle there is always a global `angular' $n$-form $\psi$
whose restriction to each fibre generates the cohomology of the
fibre. The form $\psi$ is not in general closed, the obstruction
being given by the Euler class  $e(\Lambda^+)$ (see for example
\cite{milnorstasheff} for an introduction to characteristic
classes). $\Lambda^+$ has rank 3 and therefore has an
orientation-reversing automorphism, given by $x \to -x$ on the
fibres, under which the Euler class changes sign. The Euler class
must therefore vanish. The angular form is then a cohomology class
that generates the cohomology of the fibres. The existence of such
a cohomology class for an orientable sphere bundle is precisely
the condition under which the Leray-Hirsch theorem states that the
cohomology of the bundle factorises
\be
H^*(Z) \cong H^*(M) \times H^*(S^2) \,.
\ee
It is then immediate to derive
\be
b_2(Z) = b_2(M) + 1 \quad \text{and}\quad  b_3(Z) = 2 b_1(M).
\ee
In particular, for the manifolds we are mainly interested in
\be\label{eq:M}
\begin{array}{c}
\qquad \qquad n \\
M = \C^2\#\overline{\bP^2}\; {\overbrace{ \# \cdots \#}}\;
\overline{\bP^2}
\,,
\end{array}
\ee
we have $b_2(Z) = n+1$ and $b_3(Z) = 0$.

We have defined $Z$ as the sphere bundle of unit self-dual 2 forms
or as the bundle of complex structures. There is a third useful
description: when the manifold $M$ is spin there are well-defined
spin bundles $\Sigma^\pm$, which are rank 2 smooth complex
bundles. We have the identifications $\Lambda^1 =
\Sigma^+\otimes\Sigma^-$ and $\Lambda^\pm =
\Sigma^\pm\otimes_S\Sigma^\pm$, where $\otimes_S$ denotes symmetrised
tensor product.
In this case $Z$ can also be defined as $\bbP(\Sigma^+)$, the
projectivisation of the spin bundle. This description is useful
locally even when the manifold $M$ is not spin.

\subsection{Twistorial super Calabi-Yau manifolds}\label{superCY}

In the context of the B model topological string theory one must
make the twistor space into a super Calabi-Yau
manifold in order for the string theory to be free of anomalies
\cite{Witten:2003nn}. The B model needs a nonvanishing global
holomorphic volume form. Such a form does not exist for twistor spaces
because the canonical line bundle $K=\Omega^3$ is nontrivial.
To overcome this problem in the case of $\bP^3$ \cite{Witten:2003nn},
which has $K={\mathcal{O}}(-4)$, one can consider
a fermionic rank four complex vector bundle
\be
E = {\mathcal{O}}(1)\oplus{\mathcal{O}}(1)
\oplus{\mathcal{O}}(1)\oplus{\mathcal{O}}(1) \,,
\ee
over the twistor space $\bP^3$. The total space of this bundle
is in fact the super Calabi-Yau $\bP^{3|4}$. The super Calabi-Yau
property follows from the fact that the `Berezinian line bundle'
\be\label{eq:berez}
B = K \otimes \Lambda^4 E \,,
\ee
is trivial because $\Lambda^4 E = {\mathcal{O}}(4)$. The Berezinian
line bundle is the generalisation of the canonical bundle to include the
fermionic coordinates.

LeBrun has presented a natural generalisation of this construction to
any twistor space \cite{lebruntalk}. The construction has two
prominent features, firstly that given any twistor space $Z$ it produces a
super Calabi-Yau manifold by appending a fermionic rank four vector
bundle $E$. Secondly, when restricted to any real curve in $Z$ the bundle
becomes ${\mathcal{O}}(1)\oplus{\mathcal{O}}(1)
\oplus{\mathcal{O}}(1)\oplus{\mathcal{O}}(1)$ over $\bP^1$. This
second point will probably be important for the local degrees of
freedom in four dimensions to remain those of ${\mathcal{N}}=4$ conformal
supergravity \cite{Berkovits:2004jj}. We give a proof of the second
point in Appendix \ref{jet}.

First note that
any twistor space is spin \cite{hitchin}.
Therefore the inverse square root of the
canonical bundle, $K^{-1/2}$, is well-defined. The rank four vector
bundle will be given by the 1-jets
\be\label{eq:easyjet}
E = J^1(K^{-1/2})\,.
\ee
Loosely, $J^1(L)$ for some line bundle $L$ has as fibre at each point
the equivalence classes of sections of $L$ determined by their value and first
derivatives at that point. More precisely, there is a short exact sequence
\be
0 \to \Omega^1\otimes K^{-1/2} \to J^1(K^{-1/2}) \to K^{-1/2} \to 0 \,.
\ee
This sequence does not split for twistor spaces. In Appendix \ref{jet} we show
how the short exact sequence gives the transition functions for the
jet bundle. The only result we need is that this sequence implies that
\be\label{eq:det}
\Lambda^4 E = K^{-1/2} \otimes \Lambda^3(\Omega^1\otimes K^{-1/2}) \,.
\ee
We define the Berezinian line bundle $B$ as before in (\ref{eq:berez}) but
now with $E$ given by (\ref{eq:easyjet}). 
However, from (\ref{eq:det}) we now have that
\bea
B & = & K \otimes K^{-1/2} \otimes \Lambda^3 \left(\Omega^1 \otimes
K^{-1/2} \right) \nonumber \\
 & = & K \otimes K^{-1/2} \otimes  K^{-3/2} \otimes K \nonumber \\
 & = & 1\,,
\eea
so the bundle is trivial and the total space is super Calabi-Yau.

Although we shall not use the details of the above argument here,
the possibility of a super Calabi-Yau construction for general
twistor spaces is important for the duality between conformal
gravity and B model topological strings to be tenable.

\section{Twistor spaces of K\"ahler manifolds and D1 brane
  charge}\label{kahler}

The complex geometry of the twistor space $Z$ encodes all the
information about the conformal geometry of the underlying
4-manifold $M$. We consider in this section the additional
structure enjoyed by the twistor space when $M$ is a K\"ahler
manifold \cite{lebrunbimer}. A two complex dimensional K\"ahler metric
is anti-self-dual if and only if it is scalar flat \cite{lebrunrev1}.
The extra structure will be of importance to us shortly
because spacetime foam with sufficient symmetry admits a K\"ahler
structure that is compatible with anti-self-duality.

Notice first of all that $Z$, being a sphere bundle, does not come
with a zero section. In general there is thus no canonical way of
embedding $M$ as a submanifold in $Z$. However, when there is a
K\"ahler form this provides precisely a nowhere vanishing section
of $\Lambda^+(M)$, and thus a section of the twistor fibration
$S(\Lambda^+)$. This section gives us a complex submanifold of $Z$
diffeomorphic to $M$. More precisely:

\vspace{0.2cm}
\noindent {\bf Theorem \cite{pontecorvo}:} {\it Let $\pi:Z \rightarrow
M$ be the twistor fibration of an anti-self-dual 4-manifold
$(M,[g])$. Suppose $Z$ contains a complex hypersurface $D$ that is
the image of a section of $\pi$. Let $J$ be the complex structure
on $M$ determined by $D$. Then there is a metric $h\in[g]$ which
is K\"ahler with respect to $J$ if and only if the line bundle of
the divisor $[D]+\sigma[D]$ is isomorphic to $K^{-1/2}_Z$.}

\vspace{0.2cm}
We can recover the K\"ahler form from the twistor data in the
following way. From the previous theorem, a section $s \in
H^0(Z,K^{-1/2})$ has a simple zero on $D\cup\bar D$, with $\bar D =
\sigma D$. Taking a
cover of $Z$ given by $\{ Z_1 = Z\backslash D$, $Z_2 = Z\backslash
\bar D\}$, $s$ is holomorphic and nonzero on $Z_1
\cap Z_2$ and so $r_{12} = s^{-2}$ defines a cohomology class $r
\in H^1(Z, K)$. We now recall how the Penrose transform
\cite{lebrunbimer,Witten:2003nn} maps
$r$ to a self-dual closed 2-form of type $(1,1)$ for the complex
structure defined by $D$. The Penrose transform involves a contour
integral along the twistor lines $F_x \cong \CP^1$, which are the
fibres of the twistor fibration above each point $x \in M$.
Let $\lambda^a$ be homogeneous coordinates on the $\CP^1$ fibre viewed
as $\bbP(\Sigma^+)$ and use standard spinor indices so that $v_a
\in \Sigma^+, v_{\dot a} \in \Sigma^-$. The spinor indices are raised
and lowered with $\epsilon_{ab}$ and $\epsilon_{\dot a \dot b}$. A
1-form on $M$, for example, is written as $\phi_{a \dot a}$. Since
$K|_F = \cO(-4)$ then $s_x = s|_{F_x} \in H^0(F_x, \cO(2))$ and
one can write $s_x = \omega_{ab}(x) \lambda^a \lambda^b$. Notice
therefore that $r_x \in H^1(F_x,
\cO(-4))$. The Penrose transform associates to $r_x$ a
self-dual two form via a contour integral on $F_x$:
\begin{equation}\label{eq:kahler}
k_{ab}(x) = \oint_{\Gamma \subset F_x} \lambda_c d \lambda^c \,
\lambda_a \lambda_b r_x =
\oint_{\Gamma \subset F_x} {\lambda_c d\lambda^c \, \lambda_a \lambda_b \over
(\omega_{ef}(x) \lambda^e \lambda^f)^2} = {\omega_{ab}(x) \over
({\rm det} \omega(x))^{3/2}} \, .
\end{equation}
It can be proven that $k$ is the K\"ahler form of a scalar-flat
anti-self-dual metric on $M$ \cite{lebrunbimer}.

To see that this form is of type $(1,1)$ on $M$ notice
\cite{Atiyah:wi} that for each point $\lambda \in F_x$ there is an
isomorphism $T^*_x M \simeq \Sigma^-$ given by $v^{a \dot a}
\mapsto \phi^{\dot a} = \lambda_a v^{a \dot a}$. Because $\Sigma^-$ is a
complex vector space, this map induces a complex structure on $T^*
M$ that depends on $\lambda$ up to multiplication by a number.
With this complex structure, the 1-forms $e^{\dot a} =
\lambda_b dx^{b \dot a}$ are by definition of type (1,0). We can
decompose a 1-form in its components of type (1,0) and (0,1) using
the projectors $\Pi^b{}_a =
i \bar \lambda^b \lambda_a$, $\bar \Pi^b{}_a = i \lambda^b \bar \lambda_a$,
assuming $\lambda$ is normalized so that $\lambda^a \bar \lambda_a =
-i$. Then we have $dx^{a \dot a} = \bar
\lambda^a e^{\dot a} + \lambda^a \bar e^{\dot a}$. Inserting this relation into
the expression (\ref{eq:kahler}) for the K\"ahler form, $k \propto
\omega_{ab} dx^{a \dot a} dx^{b \dot b} \epsilon_{\dot a \dot b}$ we see that the
terms of type (2,0) and (0,2) are proportional to $\omega_{ab}
\lambda^a \lambda^b$. These terms vanish exactly at $\{s_x=0\} = (D\cup \bar
D)\cap F_x$. However, this is precisely where the residue of the
contour integral (\ref{eq:kahler}) lies and hence where the form is
evaluated. It follows that
the K\"ahler form is of type (1,1) in the complex structure determined
by $D$ (or $\bar D$), as it should be.

The main observation we make is that we can now write a natural
meromorphic 3-form on $Z$ corresponding to $s^{-2}$. Locally, in a
patch where the twistor fibration can be trivialised, it is given
by
\begin{equation}\label{threeform}
\Omega =  s^{-2}\, \lambda^c d\lambda_c \, \lambda_a \lambda_b \, dx^{a \dot a}
dx^{b \dot b} \, \epsilon_{\dot a \dot b} \,.
\end{equation}
This 3-form will have quadratic singularities on a $D \cup \bar
D$. The previous discussion then shows that integrating along a
contour in a fibre gives the K\"ahler form,
\begin{equation}\label{eq:related}
\oint_{\Gamma \subset F_x}  \, \Omega = k(x) \,.
\end{equation}
It follows that we can recover the K\"ahler moduli of the
4-manifold by integrating $\Omega$ on suitably chosen 3-cycles in
$Z$
\be\label{eq:integrals}
\int_{\Sigma^{(2)}} k = \int_{\Sigma^{(3)}} \Omega
\,.
\ee
This relationship is exciting because it relates a measure of four
dimensional spacetime foam to an integral in six dimensions that
detects D1 brane charge. Such a relationship is what we had anticipated
in the introduction. It might seem contradictory that in
(\ref{eq:related}) a $(1,1)$ form in four
dimensions is related through a contour integral to a $(3,0)$-form in six
dimensions. However this is what happens as we will now illustrate for
the case of $\CP^3$.

In section \ref{blowup} we will explicitly construct the meromorphic 3-form
for a class of twistor spaces which are bimeromorphically
algebraic. Let us now, as an example, spell out this
construction for the case of $\bP^3$ (strictly $\bP^{\prime 3}$).
The corresponding 4-manifold is just $\C^2$. The divisors
$D$ and $\bar D$ will intersect on a real line which is the line that
should be removed from $\CP^3$ to get the twistor space for $\C^2$
rather than $S^4$. In fact, divisors on $\CP^3$ are hyperplane
sections and they always intersect at a line. We use here the
notation of \cite{Witten:2003nn} so that $\CP^3$ has homogeneous
coordinates $\{\mu_{\dot a},\lambda^a\}$.
The fibre over a point $x$ of $\C^2$ is then
\be
\mu_{\dot a} = x_{a \dot a} \lambda^a \,,
\ee
and removing the point at infinity means that
$\lambda^a \neq (0,0)$. It is easy to see that the 3-form
(\ref{threeform}) becomes
\be
\Omega = s^{-2} \lambda^a d\lambda_a \wedge d\mu^{\dot a} \wedge d\mu_{\dot
a}\,.
\ee
Without the prefactor $s^{-2}$ this is the natural 3-form on
$\bP^3$ with values in $\cO(4)$. The prefactor turns it into an
honest (scale-invariant) 3-form but with singularities. Another way to
obtain a well defined form is to add fermionic directions
\cite{Witten:2003nn}. We shall discuss the relationship between these
singular forms and the superforms later.
The meaning of the singularity has been discussed above; for each choice of a
section $s$ we get a K\"ahler form on $\C^2$. The involution acts
as $\sigma: \lambda^a \mapsto \bar
\lambda_a$. An example of a section we could take is
$s = \delta_{ab} \lambda^a \lambda^b$. We see
that $D = \{\lambda^1 = i \lambda^2\}$, $\bar D = \{\lambda^1 = - i
\lambda^2 \}$ and $D \cap \bar D = \{ \lambda^a=0 \}$ which is the
line at infinity that we removed in these coordinates. The K\"ahler
form computed from $s$ is then
\be\label{eq:flatkahler}
k = \delta_{ab} \epsilon_{\dot a \dot
b} dx^{a
  \dot a} \wedge dx^{b \dot b} = dx^{11} \wedge dx^{12} + dx^{21}
\wedge dx^{22} = e^{\dot{1}} \wedge \bar e^{\dot{2}} - e^{\dot{2}} \wedge \bar e^{\dot{1}}\,,
\ee
where we used the normalisation $\lambda^a \bar \lambda_a =
-i$ and defined $e^{\dot a} = \lambda_b dx^{b \dot a}$, as discussed above.
We see that (\ref{eq:flatkahler}) is indeed a $(1,1)$-form and gives a self-dual
K\"ahler form on $\C^2$.
Of course $\C^2$ does not have nontrivial 2-cycles about which we can integrate
the K\"ahler form. This corresponds to the absence of D1 brane charge
in $\CP^3$.

It is remarkable that the twistor correspondence allows us to
associate a meromorphic form in 3 complex dimensions to
an anti-self-dual K\"ahler manifold $M$ in 4 real dimensions.
This appears to set up a correspondence between a physical theory on $M$
that depends on the K\"ahler moduli and a
theory on $Z$ that depends on the complex moduli. This is
reminiscent of mirror symmetry, but with a change in
dimensions involved. On the complex side of the story we are
embedding the theory into the topological B model
on a supermanifold extension of $Z$ whilst on the K\"ahler side we are
embedding the theory into (super)conformal gravity.

The fact that we would like to identify the integral of the $(3,0)$-form
around a 3-cycle as D1 brane charge is the statement that
\cite{Aganagic:2003qj} in the presence of D1 branes, or in a space
obtained from the backreaction of D1 branes, we have
\be\label{eq:quantise}
\int_{\Sigma^{(3)}} \Omega = g_s N \,,
\ee
where $N$ is the number of D1 branes linked by $\Sigma^{(3)}$.
In the topological string context $N$ is of course an
integer. Comparing with (\ref{eq:integrals}) then suggests a tantalising
quantisation of the K\"ahler moduli in the four dimensional
theory. This seems very similar to the quantisation of K\"ahler
moduli that was found in the six dimensional spacetime foam studied in
\cite{Iqbal:2003ds}.

We noted below equation (\ref{eq:M}) that for the twistor spaces
that we are interested in, describing spacetime foam, we have
$b_3(Z)=0$. Thus the only way for the closed 3-form
$\Omega$ to admit a nonzero period is if it is singular. Such a 
singularity is related to the nontriviality of the canonical bundle. 
More concretely, we have that $d\Omega$ will be a 4-form
with delta function support on the divisor $D \cup \bar D$. One
might have expected support on a 2-cycle rather than a 4-cycle,
corresponding to the location of a brane. This is not possible
however as the singular locus of a meromorphic 3-form will always
be a complex codimension-one submanifold. Thus we should think of
the twistor space as already incorporating the backreaction of the
D1 branes.

In the following section we shall study explicitly known twistor
spaces for $\C^2$ blown up at $n$ points using the framework we have
just described. Explicit asymptotically flat anti-self-dual K\"ahler
metrics and the corresponding twistor spaces are known in the case
when the $n$ blown up points are collinear \cite{lebrun}. They are
also known to exist if the configuration of points is sufficiently
close to collinear \cite{lebrunbimer}. Therefore the spacetime foam in
conformal (super)gravity has a `K\"ahler sector' and it is this sector
which we are studying.

\section{Twistor spaces for blowups}\label{blowup}

\subsection{Complex geometry and topology of the twistor spaces}

LeBrun has explicitly constructed twistor spaces for K\"ahler
anti-self-dual metrics on $M=\C^2\# \overline{\bP^2}\; \# \cdots
\#\; \overline{\bP^2}$ \cite{lebrun,lebrunbimer}. The spaces are
not the twistor spaces of the most general conformally
anti-self-dual metric on $M$ but are rather special points in the
moduli space of anti-self-dual metrics at which all the blown up
points are collinear. These have the remarkable property that the
corresponding twistor spaces are bimeromorphically algebraic. That
is, we may start with a singular algebraic threefold and obtain
the twistor space by performing blowups and blowdowns. These
spaces will allow a concrete realisation of the ideas discussed in
previous sections and will further allow us to make a connection
with super Calabi-Yau manifolds.

The construction starts by considering a singular threefold $\widetilde{Z}$.
Take $\bP^1\times\bP^1$ with homogeneous
coordinates $[z_0,z_1]$ and $[\zeta_0,\zeta_1]$. One then considers
the total space of a projectivised bundle over $\bP^1\times\bP^1$
\be
{\mathcal{B}} \equiv
\bbP \left[ {\mathcal{O}}(n-1,1) \oplus {\mathcal{O}}(1,n-1) \oplus
  {\mathcal{O}}\right]\,,
\ee
with coordinates for the fibres $x \in {\mathcal{O}}(n-1,1)$, $y \in
{\mathcal{O}}(1,n-1)$ and
$t \in {\mathcal{O}}$. Note that ${\mathcal{B}}$ is a four complex
dimensional manifold obtained from $\C^7$ via the following three
identifications
\bea\label{eq:cstar}
\left[z_0,z_1\right] & \sim & \lambda [z_0,z_1] \,, \nonumber \\
\left[\zeta_0 ,\zeta_1\right] & \sim & \mu [\zeta_0,\zeta_1] \,, \nonumber \\
\left[x,y,t\right] & \sim & \nu [\lambda^{n-1}\mu\,
  x,\lambda\mu^{n-1}\, y, t] \,.
\eea

Now consider the singular threefold given by a hypersurface
\be
\widetilde{Z} \subset {\mathcal{B}} \,,
\ee
defined by
\be\label{eq:define}
F \equiv xy -  t^2 \prod_{j=1}^n P^j = 0 \,.
\ee
In this expression, the $P^j$ are $n$ polynomials in
$(z_0,z_1,\zeta_0,\zeta_1)$
which we now define.
Note that the manifold $\bP^1\times\bP^1$ admits an antiholomorphic involution
$\s$ given by
\be\label{eq:sigma}
\s \left([z_0,z_1], [\zeta_0,\zeta_1] \right) \mapsto
\left([\bar{\zeta_0},\bar{\zeta_1}], [\bar{z_0},\bar{z_1}] \right) \,.
\ee
In order to obtain later the twistor space for the noncompact manifold $M$
we should remove from $\bP^1\times\bP^1$ the line $S \subset
\bP^1\times\bP^1$ given by the fixed points of $\sigma$. This
will correspond to removing the point at infinity in four dimensions.
Consider $n$ $\s$-invariant curves in $\bP^1\times\bP^1 - S$, which we
denote $\{C^i\}_{i=1}^n$. These curves are specified as the zeros
of $n$ polynomials
\be
P^i \in \Gamma \left(\bP^1\times\bP^1, {\mathcal{O}}(1,1) \right) \,.
\ee
Explicitly these are
\be\label{eq:curves}
P^i = a^i_{mn} \zeta_m z_n \,, \quad \text{with} \quad a^i = a^i{}^{\dagger}\,.
\ee
We assume that the curves are nondegenerate and generic, implying that
they all mutually intersect at precisely two points. The curves are
illustrated in figure 1.
\begin{figure}[h]
\begin{center}
\epsfig{file=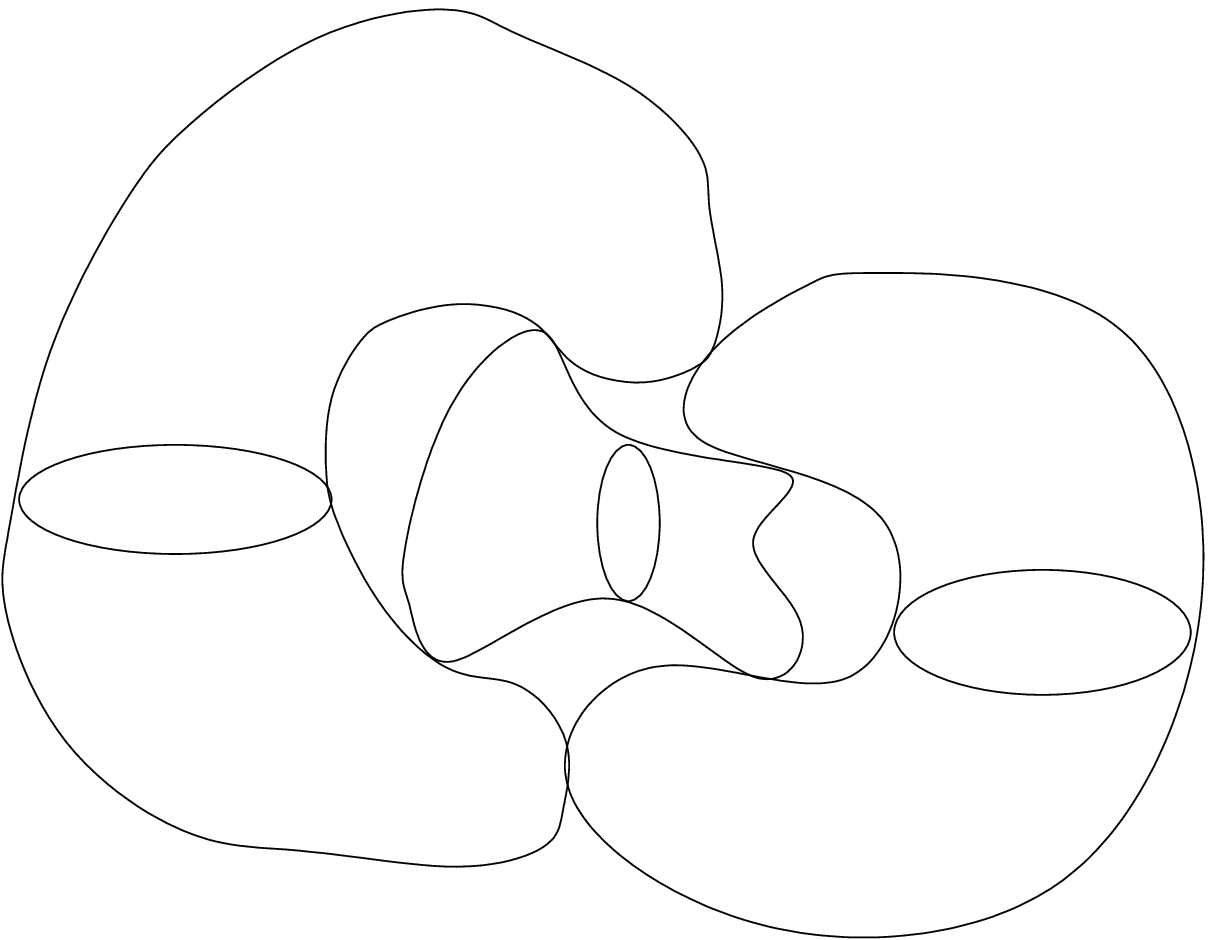,width=5cm}
\end{center}
\noindent {\bf Figure 1:} Two-cycles $C^1, C^2$ and $C^3$ mutually
intersecting at two points.
\end{figure}

The threefold $\widetilde{Z}$ is described topologically as
follows. Away from the
curves $P^i=0$ in $\bP^1\times\bP^1$ the manifold is a $\bP^1$
fibration over $\bP^1\times\bP^1$. Above the union of the curves
$X^{n} = \cup_{i=1}^{n} C^i$ the
fibration degenerates to two spheres joined at a point, which may be
written $S^2\vee S^2$. The degeneration is regular except for the
points where two curves intersect. Each pair of curves intersect at
two points, so there will be a total of $n(n-1)$ singular points.
The fibration is shown in figure 2.
\begin{figure}[h]
\begin{center}
\epsfig{file=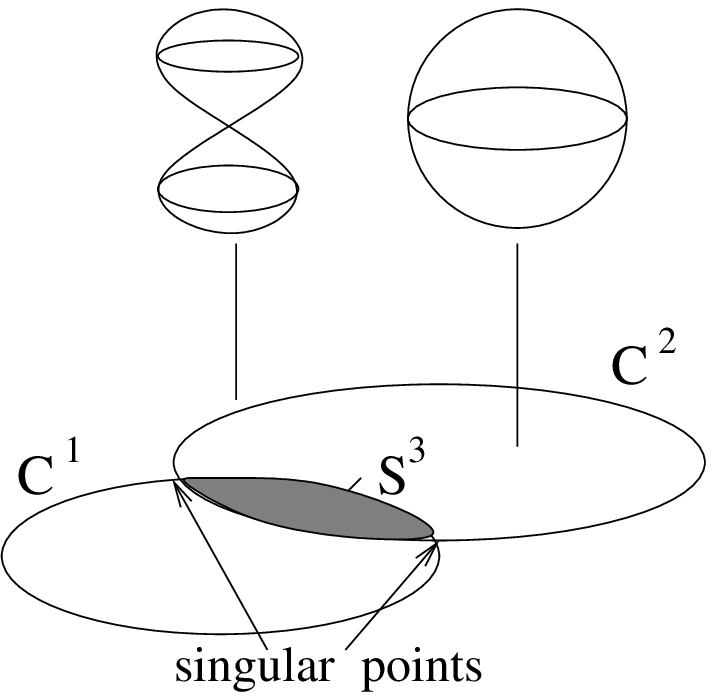,width=6cm}
\end{center}
\noindent {\bf Figure 2:} Two curves, $C^1$ and $C^2$, in $\bP^1 \times
  \bP^1$ with their singular fibration. The shaded region in the base
  corresponds to an $S^3$ in the full space.
\end{figure}

We can make a simple observation on the topology of this singular
space. Note that $S^2\vee S^2$ is formed
by collapsing an $S^1$ in $\bP^1=S^2$. Therefore every closed loop in
$X^n$ bounds a disc, $D^2$, with an $S^1$ fibration that collapses on
the boundary of the disc. This gives a homology $S^3$ in the full
space. So we have
\be\label{eq:observation}
b_3(\widetilde{Z}) = b_1(X^{n})\,.
\ee
It is easy to calculate $b_1(X^{n})$ using the Mayer-Vietoris
sequence. The computation is given in Appendix \ref{b3}. The result is
\be
b_1(X^{n}) = b_3(\widetilde{Z}) = (n-1)^2\,.
\ee

To obtain the twistor space we must resolve the singularities
of $\tilde Z$. This can be done in two ways which give the same
result: by a sequence of blowups and blowdowns or by taking
small resolutions. In the text we explain the second method, more
familiar to physicists, and discuss the equivalence with the first
method in Appendix \ref{small}.

Under the assumption of genericity that we are making, the curves
$C^i$ have only isolated intersections. We can then choose local
coordinates $\{ w_1, w_2 \}$ for $\CP^1 \times \CP^1$ centred at one of
the intersections and such that the two curves are given by the
equations $w_1=0,\, w_2=0$ respectively. The equation for the
threefold then locally reads $xy=w_1 w_2$. This has a conifold singularity
and it is well known that it can be resolved in such a way that
the singular point is replaced by a $\CP^1$. The resolved space is
(locally) the total space of a bundle $\cO(-1) \oplus \cO(-1)
\rightarrow \CP^1$. If this bundle has coordinates $\{ u,v,
[Z_1,Z_2]\}$, one defines a map
\be\label{eq:map}
(x,y,w_1,w_2) = (u Z_1, v Z_2, uZ_2, v Z_1)\,.
\ee
The map is an isomorphism away from the zero
section $u=v=0$ that is mapped to the singular point. The two
intersecting curves are $C^1 = \{x=y=w_1=0\}$ and $C^2 =
\{x=y=w_2=0\}$. In terms of the new coordinates one has $C^1=\{u=Z_2=0\}$,
$C^2=\{v=Z_1=0\}$ and therefore they have no intersection in the
resolved space. Figure 3 shows the disconnection of the cycles after
the small resolution.
\begin{figure}[h]
\begin{center}
\epsfig{file=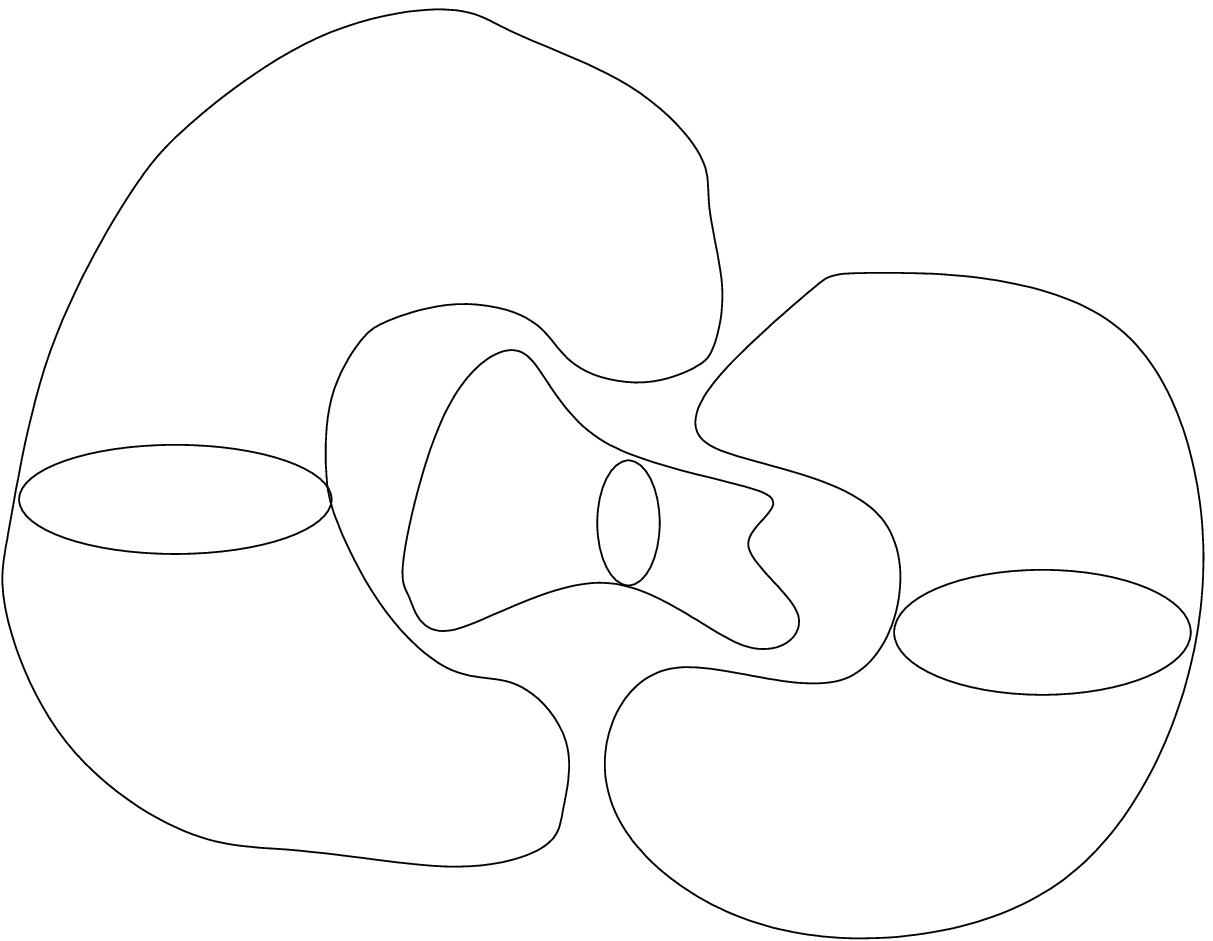,width=5cm}
\end{center}
\noindent {\bf Figure 3:} The resolution disconnects the cycles of
Figure 1. The figure is a little misleading as the separation is in
the resolved $\CP^1$ rather than in the $\{z,\zeta\}$ coordinates.
\end{figure}

The fact that the cycles $C^i$ above which the $S^2$ fibre degenerates
are no longer connected implies that the resolved space satisfies
\be
b_3(Z) = 0 \,,
\ee
in agreement with our observation below equation (\ref{eq:M}). In
Appendix \ref{trans} we 
give a more rigorous direct calculation of the Betti numbers of the
twistor space to further find
\be
b_2(Z) = n+1 \,,
\ee
again in agreement with the general considerations in section
\ref{twistsp}.

One might wonder what happens if we deform the nodal singularities
rather than taking small resolutions. These two desingularisation
are related through geometric transitions that are of general interest
in string theory. The deformed space however is not a twistor space.
In the course of calculating the Betti numbers of $Z$ in Appendix
\ref{trans} we 
consider deformation of the singularities as well as small resolutions.

There is one final technical step in the construction of the twistor
space. The singular space $\widetilde{Z}$ includes the divisors
\be\label{eq:divisors}
E_x = \{x=t=0\} \quad \text{and} \quad E_y=\{y=t=0\} \,,
\ee
that are not affected by the resolutions we have
just described. Each of these may be identified via projection to
the base with $\CP^1\times\CP^1$ and their normal bundles are
${\mathcal{O}}(-1,1-n)$ and ${\mathcal{O}}(1-n,-1)$ respectively. They
may therefore be blown down to give curves $\CP^1$ with normal bundle
${\mathcal{O}}(1-n) \oplus {\mathcal{O}}(1-n)$. Performing these two
blowdowns gives the desired twistor space $Z$ of $M$.

\subsection{The meromorphic volume form}

In this subsection we will write down the explicit
meromorphic 3-form for the twistor spaces $Z$
that is associated to the four dimensional
K\"ahler form in (\ref{threeform}). In order to do this, we have to
recall firstly the
structure of the twistor lines \cite{lebrun} and secondly the
fundamental divisor $D$ in $Z$
\cite{lebrunbimer}. We will work in terms of the coordinates
$\{x,y,t,z_0,z_1,\zeta_0,\zeta_1\}$ on $\C^7$ which are subject to the
three $\C^*$ actions (\ref{eq:cstar}) and the hypersurface condition
$F=0$ (\ref{eq:define}). Often we will work in a patch with, say, $z_1
\neq 0$ and $\zeta_1 \neq 0$. In this case we will use coordinates
$z=z_0/z_1$ and $\zeta = \zeta_0/\zeta_1$.

The generic twistor line is obtained by
taking a real curve $C$ on $\bP^1 \times \bP^1$, distinct from all the
curves $C^i$ that we considered previously. The defining equation for
$C$ is a real polynomial of degree $(1,1)$
\be
H(z,\zeta) = b_{mn} \zeta_m z_n \quad \text{with} \quad b =
b^{\dagger}\,.
\ee
Now, restriction of $P$ to $C$ is a real polynomial of degree 2$n$
with zeros at the points where $C$ intersects the $C^i$s.
We can write $P|_C = f \sigma^*(f)$ for some $f$ of degree
$n$. It turns out that the twistor lines are given by \cite{lebrun}
\be\label{eq:lines}
\{H(z,\zeta) = 0,\quad x = t e^{i
  \alpha} f,\quad y = t e^{-i \alpha} \sigma^*(f)\}\,,
\ee
with $\alpha$ fixed for a given twistor line.
If the curve $C$ happens to pass through an intersection
$C^i \cap C^j$, we take its proper transform in the resolution of the
singular space $\widetilde{Z}$.
The (proper transforms of the) curves $C^i$ themselves are also
twistor lines. The equations (\ref{eq:lines}) describe a
4-real parameter family of curves with 3 parameters from $H$ and 1 from
$\alpha$. There are also some other twistor lines that pass through $t=0$ and
are given by $\{x,y,t=0,\zeta = \bar z\}$. Notice that these lines
intersect the blown-down divisors $E_x$ and $E_y$ of (\ref{eq:divisors}).

The divisor $D$ turns out to be given as follows. Start with the
divisor $\tilde D$ in $\tilde Z$ \cite{lebrunbimer}
\be\label{eq:Dtilde}
\tilde D = \{ z=c \}\,,
\ee
in terms of the local coordinate $z$. Except for a finite number
of values of $c$, $\tilde D$ will be disjoint from the
singularities of $\tilde Z$ and therefore is unaffected by the
resolution. Thus our spaces are almost foliated by these divisors.
This last property is closely related to that of being
bimeromorphically algebraic \cite{poon}. The divisor $D$ is the
image of $\tilde D$ under the projection
\be
b: \tilde Z \to Z \,,
\ee
given by the blowdowns of the divisors $E_x$ and $E_y$
(\ref{eq:divisors}). To see the topology of $D$ \cite{lebrunbimer}
note that it inherits from $Z$ the $S^2$ fibration of Figure 2
over the $\{z=c\}$ plane that degenerates over the $n$ points
where $\{z=c\}$ intersects the curves $C^i$. The collapse of an
$S^1$ in the fibration results in $D$ being given topologically as
$\C^2$ blown up at $n$ points. Thus $D$ is homeomorphic to $M$ as
expected.

From the
expression for $\tilde D$ and the action of $\sigma$ (\ref{eq:sigma}) we have
that
\be\label{eq:DDbar}
D \cup \bar D = b \biggl( \{ z=c \}\,\cup\,\{ \zeta=\bar c \} \biggr) \,.
\ee
One can easily see now that $D$ is indeed a
section of the twistor fibration. When $z$ is fixed, the equations
for the twistor lines (\ref{eq:lines}) fix all the other
coordinates and hence $D$ intersects each twistor line at
precisely one point. 

Now let $s$ be a holomorphic section of $K_Z^{-1/2}$ with divisor $D
\cup \bar D$ as in section 3. We construct a meromorphic 3-form $\Omega_Z$ by multiplying
$s^{-2}$, which is a meromorphic section of $K_Z$, by a section of
$\Omega^3_Z \otimes K_Z^{-1}$. The latter is manifestly a trivial
line bundle and so it has a unique global section, up to a scale. 
In order to exhibit this form explicitly 
it is more convenient to work in $\widetilde{Z}$ than $Z$ because we have
global coordinates on $\widetilde{Z}$. We should bear in mind however
that $\widetilde{Z}$ is not the true twistor
space. Our strategy is to construct a meromorphic 3-form
$\Omega_{\tilde Z}$ such that 
\be
\Omega_{\widetilde{Z}} = b^*\Omega_Z \,,
\ee
where as before $b :\widetilde Z \rightarrow Z$. 
With a slight abuse of notation, we write the section $s$ using
coordinates on $\widetilde{Z}$
\be
s = (z_0 - c z_1) (\zeta_0 - \bar c \zeta_1) \,.
\ee
We may take $z = z_0/z_1$ to be a coordinate on the twistor fibre, as
$z$ is transverse to $D$. Then take $\zeta = \zeta_0/\zeta_1$
and $t$ as coordinates on $D$. We need a scale-invariant
meromorphic 3-form with quadratic poles along $\tilde D \cup \bar
{\tilde{D}}$. An expression satisfying these requirements is
\be\label{eq:Ztildeform}
\Omega_{\tilde{Z}} =
\frac{dz \wedge d\zeta \wedge dt}{t (z-c)^2(\zeta-\bar c)^2}
\,.
\ee
In the next section we will explain how this form can be obtained
using the method of Poincar\'e residue. This method will also give an
expression for the form in other coordinate charts. 
It might appear worrysome that the form $(\ref{eq:Ztildeform})$ has a
singularity at $t=0$. In fact it is to be expected: if $\Omega_Z$ is
to be singular on $D \cup \bar D$, its pull-back will be singular on
$b^{-1}(D) =\tilde{D}\cup \bar{\tilde{D}} \cup \{t=0\}$. 
This can be seen explicitly by studying the 
neighborhood of the blown-down divisor using a coordinate change
similar to that used in appendix \ref{small}. Even without doing a
computation one can argue that a meromorphic 3-form must have
singularities along a divisor, \ie a codimension 1 subvariety, but the
locus $\{t=0\}$ is blown-down by $b$ and so is not a divisor in $Z$.
Therefore $\{t=0\}$ cannot be a pole of $\Omega_Z$. 

\subsection{A holomorphic superform}

In this subsection we make an informed guess at the holomorphic
volume form on a super Calabi-Yau manifold constructed from the
twistor spaces for $M=\C^2\#\overline{\bP^2}\; \# \cdots \#\;
\overline{\bP^2}$ that we described in the previous subsections.
This superform will have the property that when we integrate out
the fermionic directions we obtain the meromorphic volume form
that we have just described.

The first step is to construct a holomorphic volume form
$\Omega_{3|4}$ on a supermanifold extension of the singular space
$\widetilde{Z}$. Recall that the bosonic part of this space was
given as the hypersurface $F=0$ in the 4-fold ${\mathcal{B}}$
(\ref{eq:define}). The supermanifold extension of $\widetilde{Z}$
will be given shortly as a hypersurface ${\mathcal{F}}=0$ in a
supermanifold extension of ${\mathcal{B}}$. From now on we shall
use $Z_S,\widetilde{Z}_S$ and ${\mathcal{B}}_S$ to refer to the
supermanifold extensions of the respective manifolds.

There is a well known way of
constructing holomorphic volume forms on hypersurfaces from a singular
holomorphic form in the ambient space, called the Poincar\'e
residue map. In our context this map takes a section of
$\Omega^{4|4}({\mathcal{B}}_S,{\mathcal{F}})$, that is $(4|4)$-forms on
${\mathcal{B}}_S$ with a simple pole along
${\mathcal{F}}=0$, to a section of $\Omega^{3|4}(\widetilde{Z}_S)$. Thus
we begin by writing down a section of
$\Omega^{4|4}({\mathcal{B}}_S,{\mathcal{F}})$.

Recall that ${\mathcal{B}}$ is constructed from three $\C^*$ actions on $\C^7$
(\ref{eq:cstar}). We can work in terms of coordinates
$\{z^i\}$ of $\C^7$. More concretely, in terms of the coordinates
introduced earlier we will have $\{z^i\} = \{x,y,t,z_0,z_1,\zeta_0,\zeta_1\}$.
If we let $k$ denote a vector generating any of the three $\C^*$ actions, then
a 4-form $\Omega_{\mathcal{B}}$ defined on ${\mathcal{B}}$
should only have legs pointing transversally
to the orbits, that is $\iota_k \Omega_{\mathcal{B}} = 0$. Further, the form
must be a sum of terms that have the same overall scaling under the
$\C^*$ actions (\ref{eq:cstar}). Starting from the holomorphic volume
form on $\C^7$:
\be
\Omega_7 = dz^1 \wedge \cdots \wedge dz^7 \,,
\ee
and given that we also want a pole along
${\mathcal{F}}=0$, the natural form to write down is
\be\label{eq:4form}
\Omega_{\mathcal{B}}= \frac{\iota_{k_\lambda} \iota_{k_\mu}
  \iota_{k_\nu} \Omega_7}{{\mathcal{F}}} \,.
\ee
In this expression the various $\iota_k$ terms denote contraction with
the generators of the three $\C^*$ actions (\ref{eq:cstar}).
The numerator turns out to be the following 4-form:
\be\label{numerator}
\iota_{k_\lambda} \iota_{k_\mu}
  \iota_{k_\nu} \Omega_7 = \, (x \, dy \, dt - y \, dx \, dt + t \, dx
\, dy) (\epsilon_{ij} z^i dz^j) (\epsilon_{kl} \zeta^k d\zeta^l) \,.
\ee
However, the form (\ref{eq:4form}) is not a 4-form on
${\mathcal{B}}$ because it is not 
invariant under the rescalings (\ref{eq:cstar})
\be\label{eq:scaling}
\Omega_{\mathcal{B}} \to \lambda^2\mu^2\nu \Omega_{\mathcal{B}}
\ee
Therefore, in the spirit of \cite{Witten:2003nn}, we will form a
supermanifold ${\mathcal{B}}_S$ by adding a four dimensional
fermionic vector bundle $E$ with local coordinates $d\eta^1, \cdots,
d\eta^4$ to ${\mathcal{B}}$. The only property we require at this
point is that the
determinant line bundle $\Lambda^4 E$ has the opposite scaling to
(\ref{eq:scaling}).
It follows that the form
\be\label{eq:append}
\Omega_{{\mathcal{B}}_S} = \Omega_{\mathcal{B}}\, d\eta^1 d\eta^2
d\eta^3 d\eta^4 \,\,
\ee
is well defined on the supermanifold ${\mathcal{B}}_S$ and is a section
of $\Omega^{4|4}({\mathcal{B}}_S,{\mathcal{F}})$.

The next step is to construct from $\Omega_{{\mathcal{B}}_S}$ a
holomorphic volume
form on the singular space $\widetilde{Z}_S$ that is given by
${\mathcal{F}}=0$ in
${\mathcal{B}}_S$ (\ref{eq:define}). The Poincar\'e residue map gives
this form to be
\be\label{eq:formonZ}
\Omega_{\widetilde{Z}_S} = \frac{\Omega_{{\mathcal{B}}_S}}{d{\mathcal{F}}} \,.
\ee
More precisely, one writes 
\be
\Omega_{{\mathcal{B}}_S} = \frac{d
  {\mathcal{F}}}{{\mathcal{F}}} \wedge 
  \eta \,,
\ee
and defines 
\be
\Omega_{\widetilde{Z}_S} \equiv  \eta|_{\widetilde{Z}_S} \,.
\ee
It is clear that the residue is a
well-defined form at all points where $d{\mathcal{F}}\neq 0$, but it
can have singularities at ${\mathcal{F}}=d{\mathcal{F}}=0$. In the
present context the singularities will be at the singular points of
$\tilde{Z_S}$.

At this point we need to discuss the expression for ${\mathcal{F}}$.
The most na\"ive thing we could do is to simply take
${\mathcal{F}}=F$, with $F$ given in (\ref{eq:define}).
However, this turns out to be somewhat
unsatisfactory. The reason we are trying to construct the holomorphic
volume form is that we would like to integrate the form over a $(3|4)$-cycle
in the supermanifold to obtain a D1 brane charge. This is the
supermanifold generalisation of (\ref{eq:quantise}). Using
${\mathcal{F}}=F$ in (\ref{eq:formonZ}) would give a form whose only
fermionic dependence is $d\eta^1 d\eta^2 d\eta^3 d\eta^4$. However, upon
Berezinian integration this form will always integrate to zero. To
get a nonzero answer we need to integrate $\eta^1 \eta^2 \eta^3 \eta^4
d\eta^1 d\eta^2 d\eta^3 d\eta^4$. The simplest way to obtain such a
term and restrict to $F=0$ on the bosonic manifold is to take
\be\label{eq:curlyF}
{\mathcal{F}} \equiv F + G \eta^1 \eta^2 \eta^3 \eta^4 \,.
\ee
Similar expressions arose in the hypersurfaces of supermanifolds
considered by \cite{Sethi:1994ch}. Note that if we don't do the
fermionic integrations then the D1 brane charge would not be a number
but would be a section of a nontrivial bundle. This would then make it
difficult to compare brane charge on different manifolds or to speak
about the number of D1 branes.

In (\ref{eq:curlyF}), $G$ must be such that the scaling of $G \eta^1
\eta^2 \eta^3
\eta^4$ under the $\C^*$ actions is equal to the scaling of $F$.
We further know from the definition of Berezinian integration that
$\eta^1 \eta^2 \eta^3 \eta^4$ must scale inversely to $d\eta^1 d\eta^2
d\eta^3 d\eta^4$. Putting these facts together requires that $G$
scales as
\be\label{eq:theform2}
G \sim F\, d\eta^1 d\eta^2 d\eta^3 d\eta^4 \sim \frac{F^2}{dz^1 \cdots
  dz^7} \,.
\ee
Guided by our findings of the previous subsection, note that a
simple expression with the correct scaling is
\be
G = \frac{F}{t s^2} \,.
\ee
That is, we take the holomophic volume form to be
\be\label{eq:superform}
\Omega_{\widetilde{Z}_S} = \frac{\iota_{k_\lambda} \iota_{k_\mu}
  \iota_{k_\nu} \Omega_7 \, d\eta^1 d\eta^2 d\eta^3 d\eta^4
}{d\left[ F+\frac{F}{t s^2}\,\eta^1 \eta^2
    \eta^3 \eta^4 \right]}\,.
\ee

It is not clear to us that this construction is unique. However,
it does appear to combine the objects that are given in a simple
and natural way to obtain a well-defined form that, as we shall
see shortly, trivialises the Berezinian line bundle. However, the
proof of the pudding is in the eating and we shall see now that
upon Berezinian integration this superform reduces to our
previously obtained meromorphic volume form on $\widetilde{Z}$.

We work in the coordinate patch $z_1 \neq 0$, $\zeta_1 \neq 0$ and
$y \neq 0$. Next, let us move $dx$ to the denominator in
(\ref{eq:superform}). This gives
\be
\Omega_{\widetilde{Z}_S} \propto \frac{z_1 \zeta_1 \, dz_0 d\zeta_0
dt d\eta^1 d\eta^2 d\eta^3 d\eta^4}{1 + \frac{1}{t s^2}
\eta^1 \eta^2 \eta^3 \eta^4} \,.
\ee
The overall normalisation is not fixed uniquely so we will not
keep track of it. If we expand the fermionic components of this
superform, we find
\be
\Omega_{\widetilde{Z}_S} = z_1 \zeta_1 \, dz_0 d\zeta_0
dt d\eta^1 d\eta^2 d\eta^3 d\eta^4 + \Omega_{\widetilde{Z}}
\eta^1 \eta^2 \eta^3 \eta^4 d\eta^1 d\eta^2 d\eta^3 d\eta^4 \,.
\ee
The first term in this expression is a global holomorphic
superform, entirely analogous to that introduced in
\cite{Witten:2003nn}, that trivialises the Berezinian line bundle
and makes the supermanifold a super Calabi-Yau manifold. The
second term will recover the form $\Omega_{\widetilde{Z}}$ of
(\ref{eq:Ztildeform}) after Berezinian integration. One can also work
in other coordinate patches, \eg when $t \neq 0$, and obtain
expressions similar to ({\ref{eq:Ztildeform}) but with $dt/t$ 
replaced by $dx/x$ or $dy/y$.

\subsection{Integrating the 3-form}

The form $\Omega_{\widetilde{Z}}$ we have constructed lives on
the singular manifold $\widetilde{Z}$ and has 
poles at the singularities where $dF=0$. As we described in section 4,
the singularities of $\widetilde{Z}$ are removed by taking small
resolutions (\ref{eq:map}). 
After the resolution
the poles in the 3-form disappear. The easiest way to see this is to use
local coordinates. The suitable coordinate change is that given in
equation (\ref{eq:map}). As noted before, the
singularities of $F=0$ are all locally of the conifold type $xy - w_1
w_2 =0$ and the 3-form given by the residue map is locally
\be 
\frac{dx \, dw_1 \, dw_2}{x} \,,
\ee
when $x \neq 0$ and similarly in the other patches. After the
resolution the form is given by
\be
dZ \, du \, dv \,,
\ee
and so is manifestly smooth. Notice that a small resolution does not introduce exceptional
divisors, so the canonical bundle does not change and the pull-back
of the form will not have additional zeros or poles. 

In this section we will integrate $\Omega_{\widetilde{Z}}$ around a
contour in the twistor lines as we described in section 3 above. The
contour will not pass through any of the singular points of
$\Omega_{\widetilde{Z}}$, so the result we get will be the same as if
we had integrated $\Omega_Z$. However, the integrand will appear to
have poles that do not exist in $\Omega_Z$. All these poles will turn
out to have zero residue.

Let us apply (\ref{eq:related}) to our meromorphic 3-from
in the patch $t\neq0$ so that $t$ may be scaled
to be a constant. The expression for the K\"ahler form becomes
\be\label{eq:kahler2}
k = \oint_{\Gamma} \frac{dz \wedge d\zeta \wedge dx}{x
(z-c)^2(\zeta-\bar c)^2} \,,
\ee
where $\Gamma$ is a curve in a twistor line (\ref{eq:lines}).
Using $z$ to parameterise the line we can see that $\zeta$ is
given by
\be\label{eq:zeta}
\zeta = - \frac{A z + B}{z + \bar{A}} \,, \qquad\text{with}\quad A
\in \C,\; B \in \R\,,
\ee
where without loss of generality we have scaled $H(z,\zeta)$ in
(\ref{eq:lines}) so that $b_{00}=1$. Also from (\ref{eq:lines}) we
have
\be\label{eq:dx}
\frac{dx}{x} = i d\alpha + \frac{df}{f} \,.
\ee
Note that we are using $\{\alpha,A,\bar{A},B\}$ as coordinates on
$D$. These may be held fixed as we do the $z$ integral.

The function $f$ may be factorised to give
\be\label{eq:f}
f = \prod_{i=1}^n f_i = \prod_{i=1}^n M_i (z - \Gamma_i) \,.
\ee
The values of $M_i$ and $\Gamma_i$ are found by requiring $f
\sigma^*(f) = P|_C$. We find
\be
\Gamma_i = \frac{B+2\text{Re}(\bar{b_i} A) - d_i+\sqrt{X}}{2(b_i-A)} \,,
\ee
where have scaled (\ref{eq:curves}) to describe $C^i$ by 
\be
z \zeta + b_i z + \bar{b_i} \zeta + d_i = 0 \,, \qquad\text{with}\quad b_i
\in \C,\; d_i \in \R\,,
\ee
and
\be
X = B^2 + d_i^2 - 4(B+d_i) \text{Re}(\bar{b_i} A) - 2 d_i B +
2\text{Re}(\bar{b_i}^2A^2) + 4d_i |A|^2 + 4 B |b_i|^2 - 2|A|^2|b_i|^2 \,.
\ee
In fact, $z=\Gamma_i$ in $C$ is a point of
intersection of $C$ with $C^i$ because $f_i=0$ on $C^i$.
The remaining coefficient in (\ref{eq:f}) is given as
\be
M_i^2 = \frac{2|A|^2+d_i-B-2\text{Re}(\bar{b_i} A)+\sqrt{X}}{2(|A|^2-B)} \,.
\ee
Note that $X$ is real and hence that $M_i^2$ will be real and positive
if $X$ is positive. This is the case that we consider.
Further note that, despite appearances to the contrary, $M_i$ does not
have a pole at $|A|^2=B$ because the numerator also vanishes at this point.
We may now write the second term in (\ref{eq:dx}) as
\be\label{eq:foverf}
\frac{df}{f} = \sum_{i=1}^n \frac{dM_i}{M_i} - \sum_{i=1}^n
\frac{d\Gamma_i}{z-\Gamma_i} \,.
\ee

We see that (\ref{eq:foverf}) introduces new poles into the $z$
integral in (\ref{eq:kahler2}). One might worry that this would allow
us to obtain different results for the K\"ahler form by integrating
around differing contours, contradicting the setup we developed in
section \ref{kahler}. However, it turns out that the residues of these
new poles is precisely zero so in fact all contours around $z=c$ give
the same answer. This provides a rather nice consistency check with the fact
that in the resolved space $Z$ the 3-form $\Omega_Z$ should only have
poles on $D\cup\bar D$.

Let us perform the integration.
The parts of the integral (\ref{eq:kahler2}) involving $\alpha$ and
$M_i$ give a succinct result
\be\label{eq:firstpart}
\frac{1}{2\pi i}\oint_{\Gamma} \frac{dz \wedge d\zeta}{
(z-c)^2(\zeta-\bar c)^2} \wedge \left[ i d\alpha + \sum_{i=1}^n
  \frac{dM_i}{M_i} \right] =
- d \left[\frac{|A|^2-B}{(B+2\text{Re}(c A)+|c|^2)^2}
  \right] \wedge \left[i d\alpha + \sum_{i=1}^n \frac{dM_i}{M_i} \right] \,.
\ee
Note that this contribution is manifestly closed. The explicit
expression for $dM_i/M_i$ is rather large and unilluminating. The
remaining terms may also be integrated
\bea\label{eq:secondpart}
\lefteqn{- \frac{1}{2\pi i} \oint_{\Gamma} \frac{dz \wedge d\zeta}{
(z-c)^2(\zeta-\bar c)^2} \wedge \sum_{i=1}^n
  \frac{d\Gamma_i}{z-\Gamma_i}} \\
& & = \sum_{i=1}^n \left(d \left[\frac{c + \bar{A}}{B+2\text{Re}(c
    A)+|c|^2} \right]
\wedge \frac{d\Gamma_i}{(c-\Gamma_i)^2}
- d \left[\frac{|A|^2-B}{(B+2\text{Re}(c A)+|c|^2)^2}
  \right] \wedge \frac{d \Gamma_i}{c-\Gamma_i}  \right) \,. \nonumber 
\eea
Once again, we see explicitly that the form is closed, as it should
be. Similarly to before, the actual expression for $d\Gamma_i$ appears to be
complicated and unhelpful. The poles in (\ref{eq:firstpart}) at
$M_i=0$ are an artifact of the coordinates we are using. From
(\ref{eq:f}) we have that $M_i=0$ implies that $f=0$ and hence from
(\ref{eq:lines}) that $x=y=0$. This is a singular point on
$\widetilde{Z}$ and (\ref{eq:kahler2}) is not valid at this point. On
the other hand, the poles in (\ref{eq:secondpart}) at $c=\Gamma_i$
turn out to be exactly what we want. It is to these poles which we
now turn.

The poles at $c=\Gamma_i$ for each
$i=1...n$ have a geometric interpretation as
follows. We noted above that $z=\Gamma_i$ on $C$ is
the location of an intersection between $C$ and $C^i$. Generically,
these points will not lie on the hypersurface $D$ given by $z=c$. We
are using the coefficients $\{A,\bar{A},B\}$ that define the curve
$C$ as
coordinates on $D$ because they are fixed for a given twistor
line. The point on $D$ corresponding to a given $\{A,\bar{A},B\}$ is the
unique point where $D$ intersects $C(A,\bar{A},B)$.
Thus $c=\Gamma_i$ means that we are at a point on $D$ that also
intersects $C^i$. In the discussion following (\ref{eq:Dtilde}) above
we noted that at these points the circle fibration in $D$
collapses. These $n$ points were then related to the $n$ blown-up
$\CP^1$s in $D$. The relationship between $C$, $C^i$ and $D$ is
illustrated in figure 4.
\begin{figure}[h]
\begin{center}
\epsfig{file=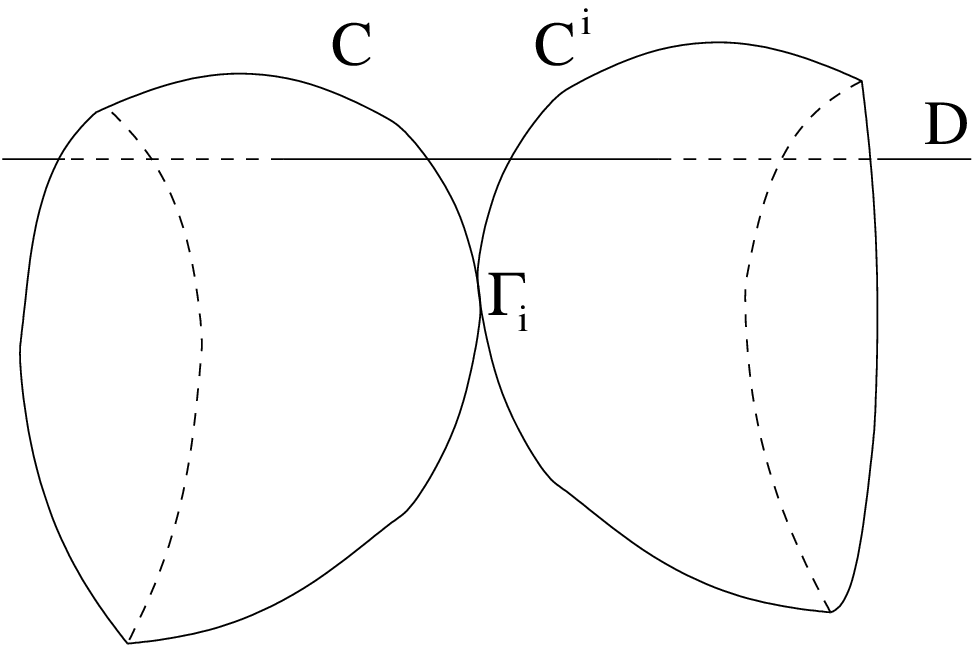,width=6cm}
\end{center}
\noindent {\bf Figure 4:} Generically $D=\{z=c\}$ does not go through
the intersection point of $C$ and $C^i$. It does so when $c=\Gamma_i$.
\end{figure}

To summarise the preceding paragraph: the K\"ahler form that we obtain
has poles associated with the $n$ blown up points of the
original self-dual manifold $M$ which we started with. This is good
because the explicit K\"ahler form for these manifolds has been written down
by LeBrun \cite{lebrun} and has precisely this property. In
principle it should be possible to find an explicit change of variables
between the 2-form we found in (\ref{eq:firstpart}) and
(\ref{eq:secondpart}) above and the form given in
\cite{lebrun}. However, this does not look particularly straightforward.
We consider the presence of the correct poles to support the
interpretation of the 3-form (\ref{eq:Ztildeform}) as the meromorphic
threeform corresponding to K\"ahler form on $M$ via the Penrose
transform.

It follows from (\ref{eq:integrals}) that when we integrate $\Omega_{Z}$ around
cycles in $Z$ given by a $\CP^1$ in $D$ and a contour in the twistor
line above each each point of $D$ we will obtain the integral of $k$
around a 2-cycle in $M$. There are $n$ such cycles because we have
blown up $\C^2$ at $n$ points. This provides a specific realisation of
the map between K\"ahler moduli for spacetime foam and D1 brane charge.

\section{Conclusions and discussion}\label{concl}

We have seen that the Penrose transform provides a natural map
between the K\"ahler moduli of K\"ahler anti-self-dual gravitational
instantons in conformal supergravity and D1 brane charge on the
corresponding twistor spaces. This provides quantitative support
for the idea that placing D1 branes in $\CP^3$ corresponds to blowing
up points in four dimensions. Such a correspondence had been
anticipated by combining three ingredients: the recent conjectures of
S-duality in topological string theory
\cite{Nekrasov:2004js,Neitzke:2004pf}, the development of twistor
string theory for $\CP^3$ \cite{Witten:2003nn} and the understanding
of spacetime foam for the A model
topological string theory on toric Calabi-Yau manifolds
\cite{Iqbal:2003ds}. The hope is that this relationship may shed light
upon spacetime foam in four dimensions or upon the role of D1 brane
instantons in topological string theory.

One interesting possibility that follows from the correspondence
between K\"ahler moduli and D1 brane charge is that the
quantisation of the number of D1 branes should translate into a quantisation
of the K\"ahler moduli of the four dimensional quantum foam. This is a
very natural expectation \cite{Iqbal:2003ds} and it would be
interesting to understand any such quantisation at a deeper level.

The appearance of the K\"ahler condition itself is also interesting. A
generic anti-self-dual gravitational instanton metric is not K\"ahler.
The correspondence suggests that the `K\"ahler sector' of spacetime
foam is particularly amenable to study using twistor string
theory. Perhaps an exact treatment of this sector is possible?

The twistor space $Z$ is not a Calabi-Yau
manifold but may always be extended to a super Calabi-Yau manifold in
a natural way. This extension implies the existence of a holomorphic
$(3|4)$-form on the super Calabi-Yau. We have suggested that the
meromorphic
3-form on $Z$ whose Penrose transform gives the K\"ahler form
could be obtained by integrating the $(3|4)$-form over the fermionic
directions. We showed explicitly how this could work for the twistor
spaces corresponding to blowups of $\C^2$. In this picture, the D1
brane charge is detected by integrating the holomorphic $(3|4)$-form over a
$(3|4)$-cycle in the super Calabi-Yau. This notion does not depend on
the K\"ahler structure of the four dimensional manifold. By using the
superforms it should be possible to extend the correspondence between
spacetime foam and D1 brane charge to general anti-self-dual
gravitational instantons.

An alternative construction of the twistor spaces for blowups that does
not emphasise the K\"ahler property is \cite{donaldsonfriedman}.
In that work, the connected sum of an anti-self-dual manifold $M$
and, say, $\overline{\CP^2}$ was formed as follows. One performs a
{\it real} blowup of a point in $M$ and a point in
$\overline{\CP^2}$. A real blowup replaces a point with an $\RP^3$.
One then glues the two copies of $\RP^3$ together to obtain the
connected sum.
In twistor space the real blowup corresponds to a complex blowup along
the $\bP^1$s corresponding to the two points. The two exceptional divisors
are then glued together and under certain conditions the resulting singular
space may be smoothed to give a new twistor space. In this case, the
corresponding four dimensional manifold $M\#\overline{\CP^2}$ is again
anti-self-dual.
Realising this construction within topological string theory
could be one way to extend the correspondence we have been studying to
general blowups.

Finally, it would be interesting to understand
the four dimensional interpretation of the D(-1) branes
in the topological B model on twistor space
\cite{Nekrasov:2004js}.
  
\section*{Acknowledgements}

We are very grateful to Ivan Smith and Maciej Dunajski for patient and helpful
explanations of various aspects of algebraic geometry and twistor
theory, respectively. We would
also like to thank Robbert Dijkgraaf, Prem Kumar,
Fabien Morel and Peter Mayr for useful discussions. S.A.H. would like to thank
the second Simons Workshop at SUNY Stony Brook for providing a
stimulating environment combining both twistors and topological
strings. S.A.H. is supported by a research fellowship from Clare
College, Cambridge. G.P. is supported by the SFB 375 of DFG.

\appendix

\section{Jet bundles}\label{jet}

In this appendix we give some technical details on the geometry of 1-jet
bundles, $J^1(L)$ (see \cite{atiyah1957}). Let $L$ be a line bundle over a
complex manifold $X$. There is an exact sequence
\begin{equation}\label{seq}
 0 \rightarrow \Omega^1 \otimes L \overset{j}{\rightarrow} J^1(L)
 \overset{p}{\rightarrow} L
 \rightarrow 0 \,.
\end{equation}
This tells us that $J^1(L)$ is an extension of $L$. Locally the
sequence is split, which means that given a cover of $X$ by $U_i$ such
that all the bundles involved are trivial on the $U_i$ and their
intersections, there are maps $h_i : L_i \to J^1(L_i)$ such that
$p \circ h_i = Id$. Together with the injection $j$, this gives
local isomorphisms $u_i^{-1}: ([\Omega^1 \otimes L] \oplus L)|_i \to
J^1(L_i)$, $(a,b) \mapsto j(a) + h_i(b)$. The inverse map is
$u_i(s) = (j^{-1}(s-h_i \circ p(s)), p(s))$. Then
\be
u_i \circ u_j^{-1} (a,b)=  (a  + j^{-1}(h_j(b) - h_i(b)),b) \,.
\ee
This shows that the class of the extension, which can be thought of
as the obstruction to a global splitting of the sequence
(\ref{seq}),  is defined by a 1-cocycle $\{h_{ij}\}$, $h_{ij}
=j^{-1}( h_j-h_i)$, with values in $\text{Hom}(L,\Omega^1 \otimes L)
\cong \Omega^1$. This cocycle is intrinsically defined as follows:
tensoring the sequence (\ref{seq}) with $L^*$ one obtains
\be
 0 \rightarrow \Omega^1  \rightarrow J^1(L) \otimes L^* \rightarrow
 \underline{\C} \rightarrow 0 \,,
\ee
where $\underline{\C}$ is the trivial line bundle. Associated to this
sequence there is a long
exact cohomology sequence with, in particular, a connecting map
$\delta : H^0(\underline{\C}) \to H^1 (\Omega^1)$. The trivial
line bundle has a global section which is just the constant
function $1$; its image under $\delta$ is the class of the
extension. In local coordinates, if $g_{ij}$ are the transition
functions of $L$ it may be shown that \cite{atiyah1957}
\be
h_{ij}= d \ln g_{ij}\,,
\ee
and therefore the extension class is just $c_1(L)$. In particular we
see that in the case of twistor spaces,
where we take $L$ to be $K^{-1/2}$, the first Chern class does not
vanish and therefore the sequence is not split, as claimed in the
text.

We can now give an explicit description of $J^1(L)$ in terms of
its transition functions $G_{ij}$. Given local trivialisations
$\phi_i:L_i
\rightarrow U_i \times \C$, and $\psi_i : \Omega^1_i \rightarrow U_i
\times \C^3$, one has $\Phi_i = ([\psi_i \otimes \phi_i] \oplus \phi_i)
\circ u_i:J^1(L_i) \rightarrow U_i \times (\C^3 \oplus \C)$
as trivialisations for $J^1(L)$. Then
\begin{equation}\label{transf}
G_{ij} = \Phi_i \Phi_j^{-1} : (v,s) \mapsto (g^{\Omega}_{ij}
g^{L}_{ij} (v) + (\psi_i h_{ij}) g^L_{ij}(s), g^L_{ij}(s)) \,.
\end{equation}
Where $g^L_{ij}$ and $g^{\Omega}_{ij}$ denote the transition functions
for $L$ and $\Omega^1$ respectively.

It is useful to observe that $J^1(L^n) \cong J^1(L) \otimes
L^{n-1}$. This can be seen by comparing two exact sequences:
one obtained from (\ref{seq}) by taking $L^n$ as the
line bundle and the other by tensoring (\ref{seq}) with
$L^{n-1}$.

The case we are interested in is $L=K_Z^{-1/2}$ for a twistor
space $Z$. We would like to compute the restriction of $J^1(L)$ to a twistor
line $C \cong \CP^1$. First, using the adjunction formula, one can see that
$K|_C \cong \cO(-4)$, so $L|_C \cong \cO(2)$. The observation
made before gives $J^1({\cal O}(2)) \cong J^1({\cal O}(1)) \otimes
{\cal O}(1)$. We also need the fact that $\Omega^1_Z|_C \cong
\Omega^1_C \oplus N^*_{C|Z} \cong \cO(-2)
\oplus \cO(-1) \oplus \cO(-1)$.
Restricting the exact sequence (\ref{seq}) to $C$, one obtains
\be\label{eq:short}
0 \rightarrow \cO(-1) \oplus \cO \oplus \cO \rightarrow
J^1(\cO(1))  \rightarrow \cO(1) \rightarrow 0 \,.
\ee
The transition function for $\cO(1)$ is $g_{ij} = z$. Using
(\ref{transf}) and noting that the extension of (\ref{eq:short})
only involves the ${\cal O}(-1)$ summand in the first term -
corresponding to 1-forms on $C$,
the other summands correspond to the normal bundle of $C$ - we can
write the nontrivial part of the transition matrix for $J^1(\cO(1))$
as
\be\label{eq:transG}
G_{ij} = \left( \begin{array}{cc} {1 \over z} & {1 \over z^2} \\ 0
& z
\end{array} \right) \,,
\ee
corresponding to a rank 2 vector bundle $E$ over $C \cong \bbP^1$. As is
well-known, every such bundle is isomorphic to a direct sum
$E = \cO(k_1) \oplus \cO(k_2)$. From (\ref{eq:transG}) we have that
$k_1 + k_2 = c_1(E) = c_1(\det E) = 1 - 1 = 0$. Now consider $E
\otimes \cO(-1) \cong \cO(k_1-1) \oplus \cO(k_2-1)$, which has transition
matrix $G_{ij} z^{-1}$. If this bundle has a global section, given in
the two charts of $\CP^1$ by vectors $(a_1(z),a_2(z))$ and $(b_1(\tilde z),
b_2(\tilde z))$, one has
\be
\left( \begin{array}{c} a_1(z) \\ a_2(z) \end{array} \right) =
\left( \begin{array}{cc} {1 \over z^2} & {1 \over z^3} \\ 0 & 1 \end{array}
\right) \, \left( \begin{array}{c} b_1(\tilde z) \\ b_2(\tilde z)
\end{array} \right) \,.
\ee
Matching the power series in $z, \tilde z$ one can see the only
solution is $b_1=b_2=0$. Therefore $E \otimes \cO(-1)$
has no global sections, which implies that $k_1,k_2 < 1$, and
therefore $k_1=k_2=0$. Thus we have that $J^1(\cO(1)) = \oplus_{i=1}^4
\cO$ which gives the result stated in the text
\be
J^1(L)|_C = \bigoplus_{i=1}^4 \cO(1) \,.
\ee

\section{Calculation of $b_3(\widetilde{Z})$}\label{b3}

In this appendix
we compute the first Betti number of
$X^{n+1} = \cup_{i=1}^{n+1} C^i$, where the $C^i$s are two-cycles
that all mutually intersect at two points. An illustration was given in
Figure 1 in the text.

Without changing the Betti numbers, we may homotopically stretch the
intersection points so that they become discs: $C^i\cap C^j=
D^2\sqcup D^2$. We may now calculate easily using part of the
Mayer-Vietoris sequence
\bea\label{eq:MV}
\cdots & \to & H_1(X_1 \cap X_2) \to H_1(X_1) \oplus H_1(X_2) \to
H_1(X) \to \nonumber \\
& \to & H_0(X_1 \cap X_2) \to H_0(X_1) \oplus H_0(X_2) \to
H_0(X) \to 0 \,.
\eea
In particular, we take $X_1 = \cup_{i=1}^n C^i$ and $X_2=C^{n+1}$.
The sequence (\ref{eq:MV}) becomes
\be
0 \to H_1(X^n)\oplus 0 \to H_1(X^{n+1}) \to \oplus_{j=1}^{2n} \Z
\to \Z\oplus\Z \to \Z \to 0\,.
\ee
The exactness of this sequence then implies that
\be\label{eq:H1}
H_1(X^{n+1}) = H_1(X^n)\oplus_{j=1}^{2n-1} \Z \,.
\ee
Taking $b_1 = \text{dim} H_1$ of (\ref{eq:H1}) and
using induction together the fact that $b_1(X^{1})=0$ one obtains
\be
b_1(X^{n}) = b_3(\widetilde{Z}) = (n-1)^2\,.
\ee

\section{Geometric transitions and topology of the twistor space}\label{trans}

This appendix calculates directly the Betti numbers of the twistor
space $Z$. In the course of the calculation we shall also consider
deformations of the singular space $\widetilde{Z}$ that are related to
the twistor space by geometric transitions.

The logic we follow is to start with a singular projective variety
$\widetilde{Z}$. The singularities are given by nodal points.
We will deform this space to a nonsingular variety
$\widetilde{Z}_{\text{def.}}$. It is straightforward to calculate
the Betti numbers of this deformed variety and then perform geometric
transitions at the deformed singular points to obtain a small
resolution of the initial singular space
$\widetilde{Z}_{\text{res.}}$. The advantage of doing this is that we
can now read off the Betti numbers of the resolved space, which would
have been harder to calculate directly. Finally, the twistor space $Z$
itself is given by blowing down two divisors in the resolved
space. Schematically:
\be\label{eq:logic}
\begin{CD}
\widetilde{Z} @>\text{deform}>> \widetilde{Z}_{\text{def.}}
@>\text{transition}>> \widetilde{Z}_{\text{res.}} @>\text{blowdown}>> Z \,.
\end{CD}
\ee

The deformed space is given by
\be\label{eq:deformed}
xy = t^2 \left[ \prod_{j=1}^n P^j + \epsilon P^{n,n}\right]\,,
\ee
where $\epsilon$ is a small parameter and $P^{n,n}$ is a generic
polynomial with homogeneity $n$ in $z_m$ and $\zeta_m$. The right hand
side of the equation for the deformed space (\ref{eq:deformed}) will
not have double roots and the resulting space will be regular.

The deformation will replace the singular points with finite $S^3$s. The
strategy is now as follows. We will first calculate
$b_3(\widetilde{Z}_{\text{def}})$ and
$b_2(\widetilde{Z}_{\text{def}})$. The geometric transition then
degenerates these cycles and replaces them with the
$\bP^1s$ that are generated by the small resolution. There is
a relationship between the Betti numbers before and after the geometric
transition, see for example \cite{smith} Theorem 2.11,
\bea\label{eq:transition}
b_3(\widetilde{Z}_{\text{res.}}) & = &
b_3(\widetilde{Z}_{\text{def.}}) - 2r \,, \nonumber \\
b_2(\widetilde{Z}_{\text{res.}}) & = &
b_2(\widetilde{Z}_{\text{def.}}) + n(n-1) -r \,,
\eea
where $n(n-1)$ is the number of nodes of the singular space
and where the degenerating three-cycles span an $r$ dimensional subset
of $H_3(\widetilde{Z}_{\text{def.}})$. More precisely
\be\label{eq:r}
r = b_3(\widetilde{Z}_{\text{def.}}) - b_3(\widetilde{Z}) \,.
\ee
The result (\ref{eq:transition}) is intuitively reasonable
\cite{smith}. For the first line in (\ref{eq:transition}) we can think
that for every three-cycle we degenerate, we lose another one
by Poincar\'e duality. For the second line we should think about how
the $r$ three-cycles that degenerate give homology relations between the
new two-cycles. A two-cycle is created at each of the $n(n-1)$
nodes. However, these can be the boundaries of three-chains that
were three-cycles before the transition.

Note that the formulae in (\ref{eq:transition}) do not apply to the
well-known conifold transition because in that case the manifolds
involved are noncompact and Poincar\'e duality is different.

We can get $b_3(\widetilde{Z}_{\text{def.}})$ using the same method as
in the singular case. The same arguments as in Appendix B imply that
\be
b_3(\widetilde{Z}_{\text{def.}}) = b_1(X^{n}_{\text{def.}}) \,,
\ee
where $X^{n}_{\text{def.}}$ is the zero space of the deformed
polynomial in (\ref{eq:deformed}). The deformation has smoothed out
the points of intersection, so
$X^{n}_{\text{def.}}$ is now just a Riemann surface. This is
illustrated in Figure 5. We can
calculate $b_1(X^{n}_{\text{def.}})$ in a similar fashion to before,
using the Mayer-Vietoris sequence or by applying the adjunction
formula to the degree $(n,n)$ polynomial. Alternatively, we can easily see
directly that $X^{n}_{\text{def.}}$ is a Riemann surface with genus
$(n-1)^2$.
Therefore we have
\be
b_3(\widetilde{Z}_{\text{def.}}) = b_1(X^{n}_{\text{def.}}) = 2(n-1)^2 \,.
\ee
\begin{figure}[h]
\begin{center}
\epsfig{file=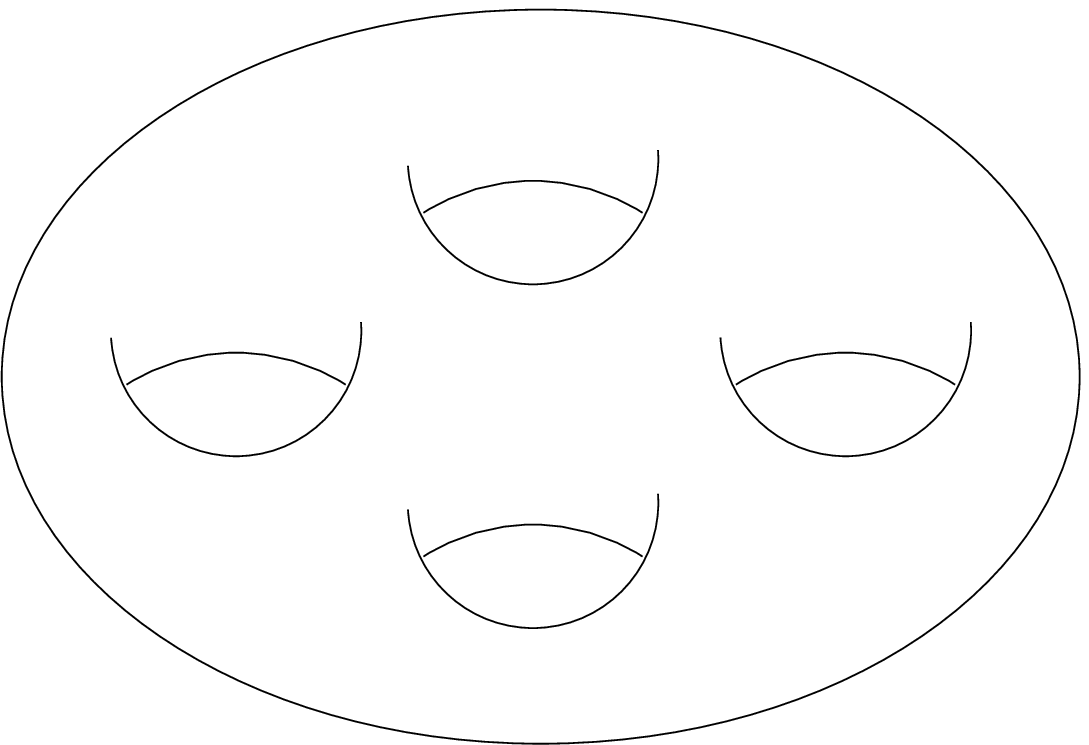,width=5cm}
\end{center}
\noindent {\bf Figure 5:} The Riemann surface resulting from smoothing
the intersections in Figure 1.
\end{figure}

\subsection{Calculation of $b_2(\widetilde{Z}_{\text{def.}})$}

In order to get $b_2(\widetilde{Z}_{\text{def.}})$, we calculate the
Euler character of the deformed space.

Recall the following two properties of the Euler character
\bea
\chi(A\cup B) & = & \chi(A)+\chi(B)-\chi(A\cap B) \,, \nonumber \\
\chi(A\times B) & = & \chi(A)\chi(B) \,.
\eea
We will write
\be
\widetilde{Z}_{\text{def.}} = A \cup B \,,
\ee
with $A$ being the degenerate fibration over $X^n_{\text{def.}}$
\be
A = X^{n}_{\text{def.}} \times (S^2\vee S^2) \,,
\ee
and $B$ the nondegenerate $\bP^1$ fibration over the
remainder of $\bP^1\times\bP^1$, see Figure 1,
\be
B = \widetilde{Z}_{\text{def.}} \setminus [X^{n}_{\text{def.}} \times (S^2\vee S^2)] = \bP^1 \times [\bP^1 \times
  \bP^1 \setminus X^{n}_{\text{def.}}] \,,
\ee
where again $S^2\vee S^2$ denotes two spheres joined at a point.

Therefore we have
\be
\chi(A) = 3\chi(X^{n}_{\text{def.}}) \,,
\ee
and
\be
\chi(B) = 2\chi(\bP^1 \times \bP^1 \setminus X^{n}_{\text{def.}}) = 2\left[4 -
  \chi(X^{n}_{\text{def.}}) \right] \,.
\ee
Putting these two results together and using the fact that $A\cap B =
\emptyset$ we have that
\be\label{eq:chi}
\chi(\widetilde{Z}_{\text{def.}}) = 8 + \chi(X^{n}_{\text{def.}}) = 10-b_1(X^{n}_{\text{def.}}).
\ee

Firstly, note that
\be
b_1(Z)=b_1(\widetilde{Z}_{\text{def.}})=b_1(\widetilde{Z}_{\text{res.}})=0
\,,
\ee
and similarly for $b_5$ by Poincar\'e duality. Note that because all
the relevant spaces are now non-singular, we may apply Poincar\'e
duality without complications. The first Betti number of the twistor
space $Z$ vanishes because $Z$ is a $\bP^1$ bundle over a simply
connected four-manifold $M$. It therefore follows from Leray's theorem
\cite{bott} that $b_1(Z)=0$. The other two spaces are related to $Z$
by blowups, blowdowns and geometric transitions which do not change the
first Betti number.

Finally, using the fact that $b_3(\widetilde{Z})=b_1(X^{n})$ and Poincar\'e
duality, it follows from (\ref{eq:chi}) that
\be
b_2(\widetilde{Z}_{\text{def.}}) = 4 \,.
\ee

\subsection{Betti numbers of the resolved space}

Now we can calculate $b_3(\widetilde{Z}_{\text{res.}})$ and
$b_2(\widetilde{Z}_{\text{res.}})$ from (\ref{eq:transition}) and
(\ref{eq:r}) to find
\be
b_3(\widetilde{Z}_{\text{res.}}) =  2(n-1)^2 - 2(n-1)^2 = 0  \,,
\ee
and
\be
b_2(\widetilde{Z}_{\text{res.}}) = 4 + n(n-1) - (n-1)^2 = n+3\,.
\ee

The last step is to go from $\widetilde{Z}_{\text{res.}}$ to $Z$ by
performing two blowdowns. The two surfaces $x=t=0$ and $y=t=0$ in
$\widetilde{Z}$ give two copies of $\bP^1\times\bP^1$. These may be
blown down \cite{lebrun} to give two $\bP^1$s as we described in the text.
The only effect of this
blowdown is to reduce the second Betti number by two
\be
b_2(Z) = b_2(\widetilde{Z}_{\text{res.}}) - 2 \,.
\ee
Therefore we obtain
\be
b_3(Z) = 0 \,, \quad b_2(Z) = n+1 \,.
\ee

It is immediately seen that this result is consistent with the cases
$n=0$ and $n=1$ where the twistor spaces are $\bP^3$ and the flag
manifold $F_3(\C)$, respectively. It is also consistent with the
general result given in section \ref{twistsp}.

\section{Small resolution vs. blowup}\label{small}

In this appendix we show how a small resolution can be obtained by
means of blowups. Since all the considerations are local we
consider, as in the text, the singular conifold in ${\Bbb{C}}^4$
\be
V = \{z_1 z_2 - z_3 z_4 = 0\}\,.
\ee
The small resolution is, as we have seen, the total space of a bundle
\be
X = \cO(-1)[u] \oplus \cO(-1)[v] \rightarrow \CP^1[Z_1,Z_2] \,,
\ee
where we have put in square brackets the coordinates for the
respective spaces.

Firstly we want to show that the blowup of $X$ along the $\CP^1$
given by the zero section $M=\{u=v=0\}$ can be identified with
the total space of a line bundle:
\be
\tilde X = \cO(-1,-1)[y] \rightarrow \CP^1[Y_1,Y_2] \times \CP^1
       [Z_1,Z_2]\,.
\ee
We give the blowup map $\pi:\tilde X \rightarrow X$ as follows:
\begin{eqnarray}
\{Y_1 \neq 0\} \,, \quad & \{u=y^{(1)}, v = {Y_2 \over Y_1} y^{(1)},
Z=Z\} \, \cr
\{Y_2 \neq 0\} \,, \quad & \{u={Y_1 \over Y_2} y^{(2)}, v = y^{(2)},
Z=Z\} \,.
\end{eqnarray}
One can see that the map is well defined, is invertible if $u \neq 0$
or $v \neq 0$,
and the inverse image of a point on $M$ is a $\CP^1$ parametrised by
$Y_1,Y_2$. This is enough to prove that $\tilde X$ is the blowup of
$X$ along $M$.

Secondly, we show that $\tilde X$ can also be interpreted as the proper
transform of the singular conifold $V$ under the blowup of
${\Bbb{C}}^4$ at the origin.
The latter is defined as
$$
\tilde {\Bbb{C}}^4 = \{(z_i) \in {\Bbb{C}}^4 , [l_i] \in \CP^3 \, | \,
z_i l_j = z_j l_i \} \,.
$$
We cover the blowup with charts $U_i = \{l_i \neq 0 \}$ and use local
coordinates $x^{(i)}$ defined by
$$ x^{(i)}_i = z_i \,, \qquad\qquad x^{(i)}_j = {l_j \over l_i} \quad, j \neq i
\,. $$
The exceptional divisor $E \simeq \CP^3$ is given in $U_i$ by
$\{x^{(i)}_i = 0\}$. The proper transform of $V$ is now obtained by
looking at its equation in local coordinates:
\begin{eqnarray}\label{proptrans}
U_1 : \quad & x^{(1)}_2 = x^{(1)}_3 x^{(1)}_4 \,, \cr
U_2 : \quad & x^{(2)}_1 = x^{(2)}_3 x^{(2)}_4 \,, \cr
U_3 : \quad & x^{(3)}_4 = x^{(3)}_1 x^{(3)}_2 \,, \cr
U_4 : \quad & x^{(4)}_3 = x^{(4)}_1 x^{(4)}_2 \,.
\end{eqnarray}
One can check that $\tilde V \cap E = \{z_i=0, l_1 l_2 - l_3 l_4 =
0\}$, which exhibits the $\CP^1 \times \CP^1$ (take $l_1 = Z_1 Y_1,
l_2 = Z_2 Y_2, l_3 = Z_1 Y_2, l_4 = Z_2 Y_1$). The remaining
coordinate parametrising $\tilde V$ is $x^{(i)}_i$, and using the
equations (\ref{proptrans})
one can see that it has the right transformation properties to be a
coordinate on the fibre of the $\cO(-1,-1)$ bundle. Thus $\tilde V =
\tilde X$ and we see that we have two routes from $V$ to the small
resolution $X$:
\be
\begin{CD}
V @>\text{small res.}>> X
@<\text{blowdown}<< \tilde{X} = \tilde{V} @<\text{proper xfm.}<< V \,.
\end{CD}
\ee

Starting from $\tilde X$ one can also obtain a different space $X'$,
isomorphic to $X$, by blowing down the other $\CP^1$, parametrised by
$Z$. The transition between $X$ and $X'$ is called a flop.

Notice that if, in the context of section \ref{blowup} in the text,
one sees the conifold as a fibration over the
$z_3,z_4$ plane which degenerates at $C^1 = \{z_3=0\}$ and $C^2 =
\{z_4=0\}$, taking the proper transforms of the $C^i$ one finds that
\begin{eqnarray*}
\tilde C^1 \cap E = & \{[l_i] = [0,0,0,1] \} = \{[Z] = [0,1],
   [Y]=[1,0]\} \, \cr
\tilde C^2 \cap E = & \{[l_i] = [0,0,1,0] \} = \{[Z] = [1,0],
   [Y]=[0,1]\} \,.
\end{eqnarray*}
Therefore the two curves remain disjoint in the resolution, regardless
of which of the
two $\CP^1$ is blown down.

\end{document}